\documentstyle{mn}
\input{psfig}
\newdimen\digitwidth
\setbox0=\hbox{\rm0}
\digitwidth=\wd0
\catcode `@=\active
\def@{\kern\digitwidth}
\def\p{\phantom{-}}
\def\g{\phantom{>}}

\begin{document}
\def\la{\mathrel{\hbox{\rlap{\hbox{\lower4pt\hbox{$\sim$}}}\hbox{$<$}}}}
\def\ga{\mathrel{\hbox{\rlap{\hbox{\lower4pt\hbox{$\sim$}}}\hbox{$>$}}}}

\title[DXRBS II. New Identifications]{THE DEEP X-RAY RADIO BLAZAR SURVEY (DXRBS) \\ 
II. New Identifications}

\author[H. Landt et al.]{Hermine Landt$^{1,2}$, Paolo Padovani$^{1,3,4}$, Eric S. Perlman$^{1,5}$, 
Paolo Giommi$^{6}$,
\newauthor Hayley Bignall$^{7}$, Anastasios Tzioumis$^{8}$ 
\\
$^1$ Space Telescope Science Institute, 3700 San Martin Drive, Baltimore, MD 21218, USA \\
$^2$ Hamburger Sternwarte, Gojenbergsweg 112, D-21029 Hamburg, Germany \\
$^3$ Affiliated to the Astrophysics Division, Space Science Department, European Space Agency \\
$^4$ On leave from Dipartimento di Fisica, II Universit\`a di Roma ``Tor Vergata'', Via della Ricerca 
Scientifica 1, I-00133 Roma, Italy \\
$^5$ Current Address: Physics Department, UMBC, 1000 Hilltop Circle, Baltimore, MD  21250, USA\\
$^6$ BeppoSAX Science Data Center, ASI, Via Corcolle 19, I-00131 Roma, Italy \\
$^7$ Department of Physics and Mathematical Physics, University of Adelaide, SA 5005, Australia\\
$^8$ Australia Telescope National Facility, CSIRO, P.O. Box 76, Epping, NSW 2121, Australia
}

\date{Accepted~~, Received~~}

\maketitle

\begin{abstract}

We have searched the archived, pointed {\it ROSAT} Position Sensitive
Proportional Counter data for blazars by correlating the 
WGACAT X-ray database with several publicly available radio catalogs,
restricting our candidate list to serendipitous X-ray sources with a
flat radio spectrum ($\alpha_{\rm r} \leq 0.70$, where $S_\nu \propto
\nu^{-\alpha}$).  This makes up the Deep X-ray Radio Blazar Survey
(DXRBS). Here we present new identifications and spectra for 106
sources, including 86 radio-loud quasars, 11 BL Lacertae objects, and
9 narrow-line radio galaxies.
Together with our previously published objects and already known
sources, our sample now contains 298 identified objects: 234
radio-loud quasars (181 flat-spectrum quasars: FSRQ [$\alpha_{\rm r}
\leq 0.50$] and 53 steep-spectrum quasars: SSRQ), 36 BL Lacs, and 28
narrow-line radio galaxies. Redshift information is available for 96\%
of these. Thus our selection technique is $\sim 90\%$
efficient at finding radio-loud quasars and BL Lacs.
Reaching 5 GHz radio fluxes $\sim 50$ mJy and $0.1-2.0$ keV X-ray fluxes 
a few $\times 10^{-14}$ erg cm$^{-2}$ s$^{-1}$, DXRBS is the faintest
and largest flat-spectrum radio sample with nearly complete ($\sim
85\%$) identification. We review the properties of the DXRBS blazar
sample, including redshift distribution and coverage of the
X-ray-radio power plane for quasars and BL Lacs.  Additionally, we
touch upon the expanded multiwavelength view of blazars provided by
DXRBS.  By sampling for the first time the faint end of the radio and
X-ray luminosity functions, this sample will allow us to investigate
the blazar phenomenon and the validity of unified schemes down to
relatively low powers.

\end{abstract}


\section{Introduction}

Blazars are the most extreme variety of active galactic nuclei (AGN)
known. Their signal properties include irregular, rapid variability;
high optical polarization; core-dominant radio morphology; flat
($\alpha_{\rm r} \la 0.5$) radio spectra; apparent superluminal
motion; and a broad continuum extending from the radio through the
gamma-rays. The broadband emission from blazars is dominated by
non-thermal processes (synchrotron and inverse-Compton radiation),
likely emitted by a relativistic jet pointed close to our line of
sight (as originally proposed by Blandford \& Rees in 1978). The properties 
of misdirected blazars are consistent with those
of radio galaxies, which form the so-called ``parent population'' within
unified schemes (e.g., Urry \& Padovani 1995).  

As a consequence of their peculiar orientation with respect to our
line of sight, blazars represent a rare class of sources, which make
up considerably less than $5\%$ of all AGN (Padovani 1997). Therefore,
currently available blazar samples suffer from small number statistics
and relatively high limiting fluxes ($\sim $ 1 Jy and a
few $\times 10^{-13}$ erg cm$^{-2}$ s$^{-1}$ in the radio and X-ray 
band respectively). The small size of these samples ($\sim 30 - 50$ objects)
implies, for example, that the derivation of the beaming parameters
based on luminosity function studies (e.g., Padovani \& Urry 1990;
Urry, Padovani \& Stickel 1991) is considerably uncertain, especially
at low powers. Moreover, our understanding of the blazar phenomenon is
mostly based on relatively bright and intrinsically luminous sources,
which means we have only sampled the tip of the iceberg of the blazar
population.

In Paper I (Perlman et al. 1998) we presented the methods, the first
85 identifications, and some preliminary results of the Deep X-ray
Radio Blazar Survey (DXRBS), a large-area ($\la 1,900$ deg$^2$
depending on the X-ray flux) survey, which reaches relatively faint
X-ray (a few $\times 10^{-14}$ erg cm$^{-2}$ s$^{-1}$) and radio
($\sim 50$ mJy, depending on declination) fluxes. The newly identified
DXRBS blazars expanded the range of $L_{\rm x}/L_{\rm r}$ values found
among blazars with emission lines by an order of magnitude, and for
the first time sampled the faint end of the radio luminosity function
of flat spectrum radio quasars (FSRQ) and BL Lacs with reasonable
statistics. Moreover, a large fraction of DXRBS BL Lacs were of the
so-called ``intermediate'' type, i.e., spanning a range of $L_{\rm
x}/L_{\rm r}$ values in between the classical low-energy peaked BL
Lacs (LBL) and high-energy peaked BL Lacs (HBL; Padovani \& Giommi
1995). This region of parameter space was almost completely unexplored
until very recently (but see also Laurent-Muehleisen et al. 1998).

This paper presents new identifications for 106 sources. In Section 2
we briefly describe the DXRBS sample and the selection technique and
discuss the radio spectral indices we obtained in the context of
source selection.  Section 3 discusses the results of our optical
spectroscopy and the makeup of the DXRBS blazar sample as of July 
2000. Section 4 reviews some of the sample properties, in particular
the redshift distribution.  Our conclusions are summarized in Section
5.

\section{The DXRBS Sample}

The selection technique and identification procedures used for DXRBS
have been described in detail in Paper I. In \S~2.1 we briefly
summarize the main points, while in \S~2.2
we discuss the derivation of the radio spectral index, the main
criterion for our candidate selection.

\subsection{Candidate Selection}

DXRBS takes advantage of the fact that all blazars are relatively
strong X-ray and radio emitters. Therefore, selecting X-ray and radio
sources with flat radio spectrum (one of the defining properties of
the blazar class) is a very efficient way of finding these rare
sources. By adopting a spectral index cut $\alpha_{\rm r} \le 0.7$
DXRBS: 1. selects all FSRQ (defined by $\alpha_{\rm r} \le 0.5$);
2. selects basically all BL Lacs; 3. excludes the large majority of
radio galaxies.

DXRBS uses a cross-correlation of all serendipitous X-ray sources in
the publicly available {\it ROSAT} database WGACAT (first revision:
White, Giommi \& Angelini 1995), having quality flag $\ge 5$ (to avoid
problematic detections), with a number of publicly available radio
catalogs. North of the celestial equator, we used the 20 cm and 6 cm
Green Bank survey catalogs NORTH20CM and GB6 (White \& Becker 1992;
Gregory et al. 1996), while south of the equator we used the 6 cm
Parkes-MIT-NRAO catalog PMN (Griffith \& Wright 1993). All sources
with radio spectral index $\alpha_{\rm r} \leq 0.7$ at a few GHz and
off the Galactic plane ($|b| > 10^{\circ}$) were selected as blazar
candidates.

For objects south of the celestial equator, where a survey at a
frequency different from the one of the PMN (6 cm) was missing when we
started this project (the NVSS [Condon et al. 1998], now available,
reaches in any case only $\delta = -40^\circ$), we conducted a
snapshot survey with the Australia Telescope Compact Array (ATCA) at
3.6 and 6 cm. This not only gave us arcsecond radio positions for our
southern sources (we use the NVSS for the northern ones) but also
radio spectral indices unaffected by variability.  In \S~2.2 we will
compare the spectral indices derived from these different surveys.

The details of the cross-correlations and the criteria for the
definition of the sample are given in Paper I. Briefly, the {\it
ROSAT} WGACAT was initially correlated with the GB6 and PMN catalogs
with a radius of $1^{\prime}$. The correlation radius was then
expanded to $1.5^{\prime}$ for the inner $30^{\prime}$ of the {\it
ROSAT} Position Sensitive Proportional Counter (PSPC) field of view,
excluding sources for which the ratio between X-ray/radio offset and
positional error was larger than 2 (and therefore likely to be
spurious). A radius of $1.5^{\prime}$ is roughly equal to twice the
combined X-ray and radio positional uncertainty of a source with PSPC
offset $\sim 30^{\prime}$ and radio flux $\sim 50$ mJy (the typical
radio flux limit of the PMN, which is declination dependent; in the
north our flux limit is given by the NORTH20 [$> 100$ mJy at 20 cm]).

In the meantime, we have somewhat expanded our criteria to increase
completeness. First, in the inner $\sim 30^{\prime}$ of the PSPC field
of view we have included flat-spectrum radio sources with ratio of
X-ray/radio offset to positional error between 2.0 and 2.5. Second, we
have added sources in the $30^{\prime} - 45^{\prime}$ PSPC offset
range having a correlation radius with the GB6 and PMN catalogs up to
$1.5^{\prime}$.

A preliminary comparison of the radio flux distributions in the inner
($0^{\prime} - 30^{\prime}$) and outer ($30^{\prime} - 45^{\prime}$)
regions of the PSPC shows no dependence on the PSPC offset for $f_{\rm
r} \ga 50$ mJy. Therefore we expect to be complete down to this radio
flux up to $45^{\prime}$. A detailed discussion of the completeness of 
DXRBS will be presented in a future paper.

\subsection{The Radio Spectral Index}

One of the strengths of DXRBS is the radio spectral index cut which
allows us to take advantage of one of the defining blazar properties,
a relatively flat radio spectrum, to select blazars very
efficiently. Ideally, the selection should be done primarily on radio
core-dominance, the ratio between core and extended radio flux, known
to be $\ga 1$ in blazars (e.g., Murphy, Browne \& Perley 1993; Perlman
\& Stocke 1993; Kollgaard et al. 1996), but derivation of this
parameter requires dedicated radio observations. DXRBS makes use of
the fact that the more core-dominated an object is, the flatter its radio
spectral index (Impey \& Tapia 1990), and sets the cut at $\alpha_{\rm
r} = 0.7$ to guarantee inclusion of all BL Lacs. (This is the limiting
value for the BL Lacs with X-ray-to-radio flux ratios typical of our
survey within the multiwavelength AGN catalog of Padovani, Giommi \&
Fiore 1997). Additionally, this cut excludes the majority of radio
galaxies.

The radio spectral indices for DXRBS sources have been derived as
follows:
\begin{itemize}
\item $\delta > 0^{\circ}$: from the GB6 and NORTH20 catalogues;\\
 $\alpha_{\rm r}$ therefore covers the $1.4 - 5$ GHz range ($6 - 20$ cm);
\item $\delta < -40^{\circ}$: from our own ATCA snapshot survey;\\
 $\alpha_{\rm r}$ therefore covers the $4.8 - 8.6$ GHz range ($3.6 - 6$ cm); 
\item $-40^{\circ} < \delta < 0^{\circ}$: in this declination range we
have access to our ATCA observations and also to PMN-NVSS data ($1.4 -
5$ GHz range). For consistency with the northern part of the survey,
we decided to use PMN-NVSS data. 
\end{itemize}

\noindent
In the following we review the ATCA observations and discuss this latter 
point.   

\subsubsection{The ATCA Observations} 
The ATCA survey was undertaken in two parts: 163 sources with $\delta
< -15^\circ$ were observed on 1995 November 11-13; and a further 108
sources (four of which were repeat observations because of bad data
during  the first run) were observed on 1997 October 27-28. During
this second run we included all sources south of $\delta=0^{\circ}$.
We observed at 4.8 and 8.6 GHz since confusion due to nearby sources 
within the primary beam is more severe 
at lower frequencies for a snapshot survey like ours. 

The ATCA data were reduced using the MIRIAD software package. During
the observations, target sources were interleaved with secondary
calibrators which were used for phase calibration. The absolute flux
density scale was determined from the primary calibrator PKS
1934$-$638. Over an entire observing run, typically 3 snapshots, each
lasting 2 minutes, were taken for every source. These snapshots were
separated in hour angle by roughly 4 hours, with some adjustments made
to allow the schedule to run smoothly.  This procedure allowed
efficient observation of a large number of objects. However, as the
ATCA is a linear array with only 6 antennae, the coverage of the
$(u,v)$ plane obtained for each source was clearly very limited.

The calibrated data were imaged, and source positions were found from
a fit to the peak pixel in the image. While these positions are
accurate to $\sim 1^{\prime \prime}$ for compact sources, the very
sparse $(u,v)$ coverage resulted in poor quality images. For an
extended source with no compact ``core,'' the position of the
brightest pixel is dependent on the sampling, hence the positional
error for an extended source may be larger than a few arcseconds.

Source flux densities were estimated directly from the calibrated
visibilities. A lower limit to the total flux density of each source
was estimated from the visibilities corresponding to the shortest
baseline (337 m and 153 m for the data taken in 1995 and 1997
respectively). For 1995, when the majority of the ATCA data were
collected, this shortest available baseline gives a synthesised beam of
FWHM $\sim 50^{\prime \prime}$ at 4.8 GHz (half of this at 8.6
GHz). Thus, for sources which have extended emission with angular
diameter much larger than this, some of the extended emission is
missing in this ATCA data.

The longest baseline in the ATCA is close to 6 km, which results in
$2^{\prime \prime}$ resolution at 4.8 GHz and $1^{\prime \prime}$
resolution at 8.6 GHz. Due to the sparse sampling of the $(u,v)$ plane
mentioned above, there are no sufficient data to fit a detailed model
to a complex source, or to make a detailed image of a source with
extended structure. The level of extended structure in each source was
estimated by comparing the visibility amplitudes corresponding to the
flux density for the shortest and longest spacings in the dataset and
confirmed by visual examination of plots of amplitude versus distance
in the $(u,v)$ plane.
 
To calculate the flux density of the unresolved, compact component, a
point source was fitted to the visibilities of each source using the
MIRIAD task {\it uvfit}.  There are two contributions to the error in
measurement of the flux density of a point source: a random noise
component of the order of 1 mJy rms, and a calibration error which is
estimated to be no larger than $2\%$ of the total source flux
density. Thus the flux density of an unresolved source is measured to
high accuracy. However, the amount of possible large-scale structure
is inherently uncertain in measurements with a sparse interferometric
array such as the ATCA. Hence, the estimated total flux density of
each source should be considered as a lower limit only.

\subsubsection{Comparison between ATCA and PMN-NVSS spectral indices} 

Besides allowing us to select candidates for spectroscopy in advance
of the completion of the NVSS, the original stated purpose of the ATCA
observations (Paper I) was to help us gauging the effect of using
non-simultaneous data to derive spectral indices, and hence to
include in our sample true blazars which might otherwise
be excluded. Our ATCA data give us radio spectral indices unaffected
by variability, since the observations at both frequencies are truly
simultaneous, but in a higher frequency range than that used for our
northern objects. On the other hand, radio spectral indices derived
from PMN-NVSS data cover the more standard $1.4 - 5$ GHz and are less
affected by either spatial filtering or poor $(u,v)$ coverage, but
involve non-simultaneous data and have angular resolution no better
than $\sim 15''$ (Condon et al. 1998). The PMN-NVSS radio spectral indices 
were 
derived by summing up all NVSS sources within $3^{\prime}$ from the
PMN position, given the different beam sizes of the two
surveys. (Padovani et al., in preparation, show that this procedure is
robust and gives radio spectral indices in very good agreement with
those derived from single dish measurements.) We expect the values of
$\alpha_{\rm ATCA}$ to be correlated with $\alpha_{\rm PMN-NVSS}$, 
with perhaps a small offset indicative of steepening in the
synchrotron spectrum at higher frequencies.

In Figure 1 we compare the values of $\alpha_{\rm ATCA}$ and $\alpha_{\rm
PMN-NVSS}$ for all sources in the declination range $-40^{\circ}
<\delta<0^{\circ}$.  The locus of equal spectral indices is
represented by a dotted line. As expected, while the scatter is fairly
large, probably due to variability and to the different spatial scales
sampled by the PMN/NVSS surveys and the ATCA (see below), the two
spectral indices are well correlated ($P > 99.99\%$).

\setcounter{figure}{0}
\begin{figure}
\centerline{\psfig{figure=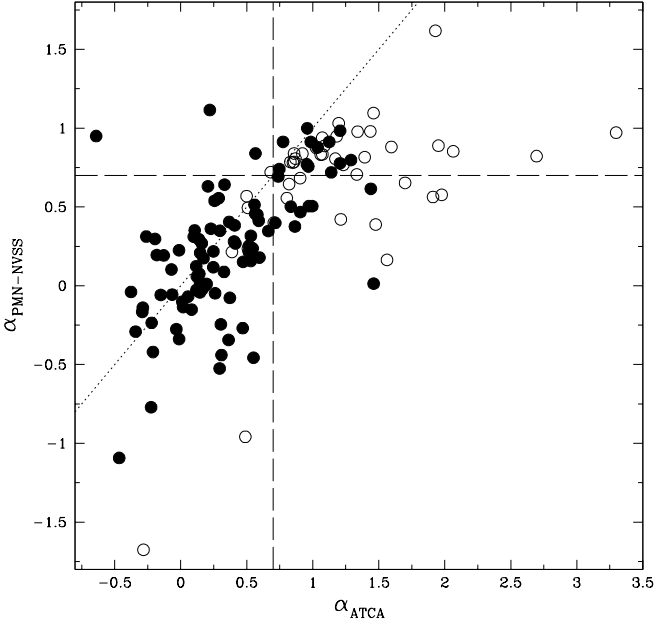,width=9cm}}
\caption{The radio spectral index derived from PMN - NVSS data ($1.4 - 5$ GHz)
compared with the radio spectral index from our ATCA observations ($4.8 - 8.6$
GHz). The dotted line represents the locus of $\alpha_{\rm PMN-NVSS} = 
\alpha_{\rm ATCA}$, while the dashed lines indicates $\alpha_{\rm r} = 0.7$
derived in each way. 
Objects with ATCA core dominance parameter at 4.8 GHz $R_{\rm 4.8} > 1$ are 
plotted as 
filled circles, while open circles indicate sources with $R_{\rm 4.8} < 1$. 
See text for discussion.}
\end{figure}

The mean values of the two spectral indices are however somewhat
different. We find $\langle \alpha_{\rm ATCA} \rangle = 0.62\pm0.06$
while $\langle \alpha_{\rm PMN-NVSS} \rangle = 0.36\pm0.04$, and a
mean difference of $\langle \Delta \alpha \rangle = \langle
\alpha_{\rm ATCA}-\alpha_{\rm PMN-NVSS} \rangle = 0.26 \pm 0.04$. This
comparison includes all sources and is therefore not appropriate: an
extended source, which will typically have a relatively steep
$\alpha_{\rm PMN-NVSS}$, will be resolved out more at 8.6 GHz than at
4.8 GHz with the ATCA, due to the smaller primary beam area at the
higher frequency. Such a source will therefore have an even steeper
$\alpha_{\rm ATCA}$. A more meaningful comparison is that between the
two spectral indices for relatively compact sources, which are the
ones we are interested in. For this purpose, we estimated a
core-dominance parameter, i.e. the ratio between core and extended
flux, from our ATCA data at 4.8 GHz. Note that only a lower limit is
available for the extended flux and therefore our core-dominance
parameters are upper limits to the true value.

Keeping all this in mind, Figure 1 shows the distribution of the two
spectral indices for sources with core-dominance parameter at 4.8 GHz
$R_{\rm 4.8} > 1$ (filled points) and $R_{\rm 4.8} < 1$ (open
points). As expected, most of the lobe-dominated ($R_{\rm 4.8} < 1$)
sources have both $\alpha_{\rm ATCA}$ and $\alpha_{\rm PMN-NVSS} > 0.5
- 0.7$. The difference between the two spectral indices is clearly
reduced for core-dominated objects. We find $\langle \alpha_{\rm ATCA}
\rangle = 0.36\pm0.05$, while $\langle \alpha_{\rm PMN-NVSS} \rangle =
0.23\pm0.04$, and a mean difference of $\langle \Delta \alpha \rangle
= 0.13 \pm 0.04$.

In the case of a direct proportionality between the two spectral
indices, all sources should populate the two areas of the diagram
defined by $\alpha_{\rm ATCA}< 0.7$, $\alpha_{\rm PMN-NVSS} < 0.7$ and
$\alpha_{\rm ATCA}> 0.7$, $\alpha_{\rm PMN-NVSS} > 0.7$ (the
bottom-left and top-right areas, denoted in Figure 1 by the two dashed
lines). This is ``almost'' the case: only 23/138 ($17\%$) of the
sources fall outside these regions. By choosing as our selection
criterion for inclusion in the sample $\alpha_{\rm PMN-NVSS} < 0.7$,
we are conservatively including most (19/23) of these outliers. Ten 
of these sources, however, have $R_{\rm 4.8} < 1$ and their blazar
classification might be questionable. As regards the four sources in
the top-left part of the figure, excluded by our criterion, one has
$R_{\rm 4.8} < 1$, while the maps of the other three show that they
are at least partially extended.

Some of the sources in Figure 1 with flat $\alpha_{\rm PMN-NVSS}$ but
marked ($\ga 0.6$) steepening at high frequencies could be GigaHertz
peaked spectrum (GPS) sources (e.g., O'Dea 1998). For these sources,
and for those with $\alpha_{\rm ATCA}> 0.7$, we give the $\alpha_{\rm
ATCA}$ values as footnotes to Table 5.

Thus our conclusion is that among our sources we observe a steepening
in spectral index between $1.4 - 5.0$ GHz and $4.8 - 8.6$ GHz of
$\langle \Delta \alpha \rangle \sim 0.25$, which reduces to 0.13 when
considering only the sources likely to be compact. By applying the
same cut of $\alpha_{\rm r}=0.7$ in the declination range $\delta <
-40^{\circ}$, where we have only $4.8 - 8.6$ GHz spectral indices
available for our candidate selection, we are missing no more than three
blazar candidates, included otherwise if $\alpha_{\rm PMN-NVSS}$ were 
available and less than $0.5$ (as the majority of our blazars are FSRQ). We 
estimate this number by considering the sources included in Fig. 1 with
$\alpha_{\rm ATCA} > 0.7$. Of these, only $7\%$ have $\alpha_{\rm PMN-NVSS} <
0.5$ and $R_{\rm 4.8} > 1$. Therefore, the use of higher
resolution data in a slightly different frequency range does not affect
strongly our selection of blazar candidates. 

As discussed in \S~3.2, spectral indices over a relatively limited
frequency range are used to establish the blazar nature of a
source. Are all our quasars with $\alpha_{\rm r} \le 0.5$ blazars? Are
they all core-dominated? We plan to address these questions in our
future work with the help of our ATCA maps, and dedicated, high resolution 
radio observations. 

\section{The Identifications}

The DXRBS sample now comprises about 350 candidates of which $\sim
85\%$ are identified.  In Paper I, classifications for 85 newly
identified objects and 97 previously known sources were given. In this
section we present 102 additional new identifications and 4 higher
signal-to-noise ratio (SNR) spectra for objects previously published
in Paper I, namely WGAJ0421.5+1433, WGAJ0528.5$-$5820, WGA
J1057.6$-$7724 and WGAJ1222.6+2934. Note that, as stated in Paper I,
the newly identified objects include not only objects meeting the strict 
selection criteria but also some lower priority sources (DXRBS objects
at low Galactic latitude, below the radio completeness limits, or with
$\alpha_{\rm r} > 0.7$). These are marked with an asterisk in Table 5.

\subsection{Identifying Candidates for Optical Spectroscopy}

Accurate ($\la 3^{\prime\prime}$) positions to pinpoint the optical
counterpart and allow spectroscopical identification were obtained
from the NVSS or from our ATCA observations. Finders were produced
using the Digitized Sky Survey (DSS) via SkyCat (Albrecht et
al. 1997). Magnitudes were derived for relatively bright objects ($B
\la 22$, $R \la 20$) from the Cambridge APM project (Irwin, Maddox \&
McMahon 1994) for the northern sources and Edinburgh COSMOS catalogue
for the southern objects (Drinkwater, Barnes \& Ellison
1995). Candidates without counterparts on DSS plates were imaged at
the KPNO 0.9 m, WIYN 3.5 m, CTIO 0.9 m, and ESO 2.2 and 3.6 m
telescopes.

Spectroscopic observations were conducted at the KPNO 2.1 m and 4 m,
and at the ESO 2.2 m and 3.6 m telescopes. Multiple grisms were used
in some runs in order to maximize SNR for faint objects and obtain
higher dispersion for brighter objects. In Table 1 we list the
resolution and approximate wavelength range of the grisms used in our
telescope runs.  In Table 2 we list the details of the
observations. Since we expected most of our sources to be quasars, the
exposure times were calculated based on this assumption. This rendered
the identification of weak emission and absorption features in our
spectra rather difficult. The spectra acquired during the September
1998 and February 1999 runs have a fairly low SNR due to bad weather
conditions. Spectra observed with the ESO 3.6 m and KPNO 4 m were
taken at parallactic angle, except in cases where the radio/X-ray
error circle contained two candidates.  Observations at parallactic
angle were not possible at the ESO 2.2 m and KPNO 2.1 m telescopes,
which require a manual slit rotation.

\vspace{0.2cm}

\begin{table}
\begin{center}
{\bf Table 1.} Grism Properties \\[0.1cm]
\nobreak 
\begin{tabular}{lcc} 
\hline 
\multicolumn{1}{c}{Telescope} & Dispersion & Range \\
& (\AA/pix) & (\AA) \\
\hline
KPNO 2.1 m &  @4.6 & 4000$-$10000\\
KPNO 4 m & @9.1 & 4300$-$10000 \\
& @4.3 & 4300$-$@8500\\
& @4.3 & 4800$-$@9500 \\ 
ESO 2.2 m & @8.4 & 3400$-$@9200 \\
ESO 3.6 m & @6.3 & 3185$-$10940 \\
& 25.0 & 3600$-$11000 \\
& @6.3 & 3740$-$@6950 \\
& @2.1 & 3860$-$@8070 \\
\hline
\end{tabular}
\end{center}
\end{table}

\vspace{0.2cm}

\begin{table*}
\begin{minipage}{150mm}
{\bf Table 2.} Log of Observations\\[0.1cm]
\begin{tabular}{lllrlllllr}
\hline 
Name & Observatory & Date & \multicolumn{1}{c}{Exp} & && Name & Observatory & Date & \multicolumn{1}{c}{Exp}\\
& & & \multicolumn{1}{c}{[sec]} & &&& & & \multicolumn{1}{c}{[sec]}\\ 
\hline 
WGAJ0010.5$-$3027 & ESO 3.6 m  & 1998 Sep & 900  &&& WGAJ1320.4$+$0140 & ESO 3.6 m  & 1999 Mar & 1200 \\       
WGAJ0014.5$-$3059 & ESO 3.6 m  & 1998 Sep & 1200 &&& WGAJ1323.8$-$3653 & ESO 3.6 m  & 1998 Feb & 2400 \\       
WGAJ0034.4$-$2133 & ESO 3.6 m  & 1998 Sep & 3600 &&& WGAJ1329.0$+$5009 & KPNO 4 m   & 1999 Feb & 1800 \\       
WGAJ0106.7$-$1034 & KPNO 2.1 m & 1998 Jun & 1200 &&& WGAJ1332.7$+$4722 & KPNO 2.1 m & 1998 Jun & 2400 \\       
WGAJ0126.2$-$0500 & ESO 3.6 m  & 1998 Sep & 900  &&& WGAJ1333.1$-$3323 & ESO 3.6 m  & 1999 Mar & 1800 \\       
WGAJ0227.5$-$0847 & ESO 3.6 m  & 1998 Sep & 900  &&& WGAJ1337.2$-$1319 & ESO 2.2 m  & 1997 May & 1500 \\       
WGAJ0251.9$-$2051 & ESO 3.6 m  & 1999 Mar & 600  &&& WGAJ1353.2$-$4720 & ESO 3.6 m  & 1999 Mar & 1200 \\       
WGAJ0305.3$-$2420 & KPNO 4 m   & 1999 Feb & 600  &&& WGAJ1359.6$+$4010 & KPNO 2.1 m & 1998 Jun & 1200 \\       
WGAJ0307.7$-$4717 & ESO 3.6 m  & 1999 Mar & 2700 &&& WGAJ1400.7$+$0425 & ESO 3.6 m  & 1998 Feb & 900 \\        
WGAJ0322.1$-$5205 & ESO 3.6 m  & 1998 Feb & 300  &&& WGAJ1402.7$-$3334 & ESO 3.6 m  & 1998 Feb & 1800 \\       
WGAJ0322.6$-$1335 & ESO 3.6 m  & 1998 Feb & 2400 &&& WGAJ1404.2$+$3413 & KPNO 4 m   & 1999 Feb & 1200 \\       
WGAJ0421.5$+$1433 & ESO 3.6 m  & 1998 Feb & 1200 &&& WGAJ1406.9$+$3433 & KPNO 2.1 m & 1998 Jun & 1200 \\       
WGAJ0427.2$-$0756 & ESO 3.6 m  & 1998 Feb & 1200 &&& WGAJ1416.4$+$1242 & ESO 2.2 m  & 1997 May & 900 \\        
WGAJ0438.7$-$4727 & ESO 3.6 m  & 1999 Mar & 2700 &&& WGAJ1417.5$+$2645 & ESO 2.2 m  & 1997 May & 1800 \\       
WGAJ0513.8$+$0156 & ESO 3.6 m  & 1999 Mar & 600  &&& WGAJ1419.1$+$0603 & ESO 2.2 m  & 1997 May & 1800 \\       
WGAJ0528.5$-$5820 & ESO 3.6 m  & 1999 Mar & 900  &&& WGAJ1420.6$+$0650 & ESO 2.2 m  & 1997 May & 900 \\        
WGAJ0533.7$-$5817 & ESO 3.6 m  & 1999 Mar & 900  &&& WGAJ1423.3$+$4830 & KPNO 2.1 m & 1998 Jun & 2400 \\       
WGAJ0534.9$-$6439 & ESO 3.6 m  & 1999 Mar & 900  &&& WGAJ1427.9$+$3247 & KPNO 4 m   & 1999 Feb & 1200 \\       
WGAJ0646.8$+$6807 & KPNO 4 m   & 1999 Feb & 2400 &&& WGAJ1442.3$+$5236 & KPNO 2.1 m & 1998 Jun & 2700 \\       
WGAJ0651.9$+$6955 & KPNO 4 m   & 1999 Feb & 1800 &&& WGAJ1457.7$-$2818 & ESO 2.2 m  & 1997 May & 1200 \\       
WGAJ0741.7$-$5304 & ESO 3.6 m  & 1999 Mar & 3600 &&& WGAJ1457.9$-$2124 & ESO 3.6 m  & 1998 Feb & 1200 \\       
WGAJ0747.0$-$6744 & ESO 3.6 m  & 1999 Mar & 1800 &&& WGAJ1506.6$-$4008 & ESO 2.2 m  & 1997 May & 900 \\        
WGAJ0829.5$+$0858 & ESO 3.6 m  & 1998 Feb & 1800 &&& WGAJ1507.9$+$6214 & KPNO 4 m   & 1999 Feb & 1200 \\       
WGAJ0847.2$+$1133 & ESO 2.2 m  & 1997 May & 900  &&& WGAJ1509.5$-$4340 & ESO 2.2 m  & 1997 May & 1800 \\       
WGAJ0853.0$+$2004 & ESO 3.6 m  & 1998 Feb & 300  &&& WGAJ1539.1$-$0658 & ESO 3.6 m  & 1998 Feb & 900 \\        
WGAJ0908.2$+$5031 & KPNO 4 m   & 1999 Feb & 1200 &&& WGAJ1543.6$+$1847 & ESO 2.2 m  & 1997 May & 1200 \\       
WGAJ0927.7$-$0900 & KPNO 4 m   & 1999 Feb & 1800 &&& WGAJ1606.0$+$2031 & KPNO 2.1 m & 1998 Jun & 1800 \\       
WGAJ0931.9$+$5533 & KPNO 4 m   & 1999 Feb & 1200 &&& WGAJ1610.3$-$3958 & ESO 2.2 m  & 1997 May & 1500 \\       
WGAJ0937.1$+$5008 & KPNO 4 m   & 1999 Feb & 600  &&& WGAJ1626.6$+$5809 & KPNO 2.1 m & 1998 Jun & 1200 \\       
WGAJ1006.1$+$3236 & ESO 2.2 m  & 1997 May & 900  &&& WGAJ1629.7$+$2117 & ESO 2.2 m  & 1997 May & 1800 \\       
WGAJ1006.5$+$0509 & ESO 2.2 m  & 1997 May & 1800 &&& WGAJ1648.4$+$4104 & KPNO 2.1 m & 1998 Jun & 2700 \\       
WGAJ1010.8$-$0201 & ESO 3.6 m  & 1998 Feb & 900  &&& WGAJ1656.6$+$5321 & KPNO 4 m   & 1999 Feb & 900 \\        
WGAJ1026.4$+$6746 & KPNO 4 m   & 1999 Feb & 1200 &&& WGAJ1656.8$+$6012 & KPNO 2.1 m & 1998 Jun & 2400 \\       
WGAJ1028.5$-$0236 & ESO 3.6 m  & 1998 Feb & 1200 &&& WGAJ1722.3$+$3103 & KPNO 2.1 m & 1998 Jun & 2400 \\       
WGAJ1028.6$-$0336 & ESO 3.6 m  & 1999 Mar & 1200 &&& WGAJ1738.6$-$5333 & ESO 2.2 m  & 1997 May & 1200 \\       
WGAJ1056.9$-$7649 & ESO 2.2 m  & 1997 May & 2100 &&& WGAJ1804.7$+$1755 & ESO 3.6 m  & 1998 Sep & 600 \\        
WGAJ1057.6$-$7724 & ESO 3.6 m  & 1999 Mar & 1200 &&& WGAJ1808.2$-$5011 & ESO 3.6 m  & 1999 Mar & 1800 \\       
WGAJ1101.8$+$6241 & KPNO 4 m   & 1999 Feb & 1200 &&& WGAJ1826.1$-$3650 & ESO 3.6 m  & 1998 Sep & 3600 \\       
WGAJ1105.3$-$1813 & ESO 2.2 m  & 1997 May & 1200 &&& WGAJ1827.1$-$4533 & ESO 3.6 m  & 1998 Sep & 1200 \\       
WGAJ1116.1$+$0828 & ESO 2.2 m  & 1997 May & 1200 &&& WGAJ1834.4$-$5856 & ESO 3.6 m  & 1999 Mar & 1800 \\       
WGAJ1120.4$+$5855 & KPNO 4 m   & 1999 Feb & 600  &&& WGAJ1837.7$-$5848 & ESO 2.2 m  & 1997 May & 1800 \\       
WGAJ1204.2$-$0710 & KPNO 2.1 m & 1998 Jun & 1200 &&& WGAJ1840.9$+$5452 & KPNO 2.1 m & 1998 Jun & 2700 \\       
WGAJ1206.2$+$2823 & ESO 2.2 m  & 1997 May & 1200 &&& WGAJ1911.8$-$2102 & ESO 3.6 m  & 1999 Mar & 1800 \\       
WGAJ1213.0$+$3248 & ESO 2.2 m  & 1997 May & 1200 &&& WGAJ1936.8$-$4719 & ESO 3.6 m  & 2000 Aug & 1200 \\       
WGAJ1213.2$+$1443 & ESO 2.2 m  & 1997 May & 1200 &&& WGAJ2109.7$-$1332 & KPNO 2.1 m & 1998 Jun & 1800 \\       
WGAJ1217.1$+$2925 & ESO 2.2 m  & 1997 May & 1200 &&& WGAJ2151.3$-$4233 & ESO 2.2 m  & 1997 May & 900 \\        
WGAJ1222.6$+$2934 & KPNO 4 m   & 1999 Feb & 1200 &&& WGAJ2154.1$-$1501 & ESO 2.2 m  & 1997 May & 600 \\        
WGAJ1223.9$+$0650 & ESO 2.2 m  & 1997 May & 1800 &&& WGAJ2159.3$-$1500 & ESO 2.2 m  & 1997 May & 600 \\        
WGAJ1225.5$+$0715 & ESO 3.6 m  & 1998 Feb & 1200 &&& WGAJ2201.6$-$5646 & ESO 2.2 m  & 1997 May & 1200 \\       
WGAJ1229.5$+$2711 & KPNO 4 m   & 1999 Feb & 1200 &&& WGAJ2239.7$-$0631 & ESO 3.6 m  & 1999 Aug & 600 \\        
WGAJ1311.3$-$0521 & ESO 3.6 m  & 1999 Mar & 900  &&& WGAJ2320.6$+$0032 & ESO 3.6 m  & 1999 Aug & 900 \\        
WGAJ1314.0$-$3304 & ESO 2.2 m  & 1997 May & 900  &&& WGAJ2329.0$+$0834 & ESO 3.6 m  & 1998 Sep & 1800 \\       
WGAJ1315.1$+$2841 & KPNO 4 m   & 1999 Feb & 2700 &&& WGAJ2330.6$-$3724 & ESO 3.6 m  & 2000 Aug & 900 \\        
\hline
\end{tabular}
\end{minipage}
\end{table*}

The acquired spectra were reduced using standard IRAF routines. The data were
bias-subtracted and flatfielded using programs in the IRAF package {\it
noao.imred.ccdred}, and the spectra were extracted, wavelength and
flux-calibrated using programs in the package {\it noao.twodspec}. Cosmic rays
were removed in the 1 and 2-dimensional data by hand. A dereddening correction
was applied to the data using the IRAF routine {\it noao.onedspec.dered} and
assuming Galactic values of extinction derived from 21-cm measurements (Stark
et al. 1992).

\vspace{0.2cm}

\begin{table}
\begin{center}
{\bf Table 3.} X-ray Positional Errors\\[0.1cm] 
\begin{tabular}{cc} 
\hline 
\multicolumn{1}{c}{PSPC} & WGACAT \\
\multicolumn{1}{c}{Center Offset} & Positional Error \\
\hline
$@0-10'$ & $13.0''$ \\
$10-20'$ & $18.1''$ \\ 
$20-30'$ & $28.6''$ \\
$30-40'$ & $36.1''$ \\
$40-50'$ & $42.0''$ \\
$50-60'$ & $53.4''$ \\
\hline
\end{tabular}
\end{center}
\end{table}

\vspace{0.2cm}

Positional information for all sources for which we announce new or
refined identifications herein are given in Table 4, including
information from WGACAT and the PMN or GB6 surveys. Column (1) gives
the WGA name of the source, columns (2) and (3) give the WGA position,
column (4) gives the source offset from the center of the PSPC field
of view, in arcmin, columns (5) and (6) give the PMN or GB6 position,
columns (7) and (8) give the position of the optical counterpart
(which is taken as the position of the corresponding NVSS or ATCA
source if within $\sim 3^{\prime\prime}$; otherwise the optical
position was measured on the finder and a footnote is given). Finally,
column (9) gives the offset between the WGA (X-ray) and optical
position, in arcsec, while column (10) gives the ratio between the
latter offset (column 9) and the sum in quadrature of the X-ray and
optical (radio) positional error. The value in column (10) corresponds
roughly to the significance in $\sigma$'s of the X-ray/radio mismatch
(e.g., a value of 2.0 implies a 95\% probability that the X-ray/radio
match is random; see Paper I for details).  The chance radio/optical
identification is very small.  Using a 3$^{\prime\prime}$ radio
positional uncertainty, the number of previously unidentified sources
in DXRBS ($\sim 250$), and a background density of stellar objects and
galaxies from faint UK Schmidt plates of 2.1 stars and 1.2 galaxies
per arcmin$^2$ (e.g., Jauncey et al. 1982), one obtains $\sim 6$
chance coincidences (or $\sim 2\%$).
 
A number of DXRBS sources were serendipitously observed by {\it ROSAT} on
more than one occasion; for completeness, we give WGACAT positions for
all observations. The WGACAT positional error is not stated
individually but is a function of the offset from the center of the
PSPC, i.e. column (4), and can be derived from this value using Table
3 (see also Paper I).

\subsection{Source Classification}

The original definition of the blazar class (cf. Angel \& Stockman
1980) emphasized the dominance of a highly polarized, variable,
nonthermal continuum over other properties. The requirements for the
classification as a blazar have nevertheless changed over the years
(see Paper I). A flat radio spectrum, which is a direct indicator of
radio core dominance (Impey \& Tapia 1990), and an optical spectrum
dominated by a non-thermal continuum and showing strong, broad
emission lines or very weak features (see below), are currently the
properties an identification as a blazar is based upon.

The blazar class is split into two subclasses: the flat spectrum radio
quasars (FSRQ) and the BL Lacertae objects (BL Lacs). Both are
believed to be radio galaxies with their jets viewed at small angles
with respect to the line of sight, with BL Lacs and FSRQ being the
``beamed'' version of the Fanaroff-Riley (FR: Fanaroff \& Riley 1974)
type I and II radio galaxies (low- and high-radio power) respectively
(e.g., Urry \& Padovani 1995).

As the optical spectrum of a blazar is one of the main observational
bases for its classification, it is important to understand its
makeup. Three components seem to be present: synchrotron emission
coming from the jet; broad and narrow lines emitted from clouds of
different densities and temperatures orbiting the central engine of
the active galaxy; and thermal stellar emission and absorption from
the AGN host galaxy. Which of these components is actually observed in
the spectrum of a radio-loud AGN appears to depend on the inclination
of the jet with respect to the observer's line of sight.

If the jet forms a relatively large angle with the observer's line of
sight, the broad line region (BLR), which is situated within a few
parsecs of the central black hole, is thought to be obscured by the
dusty torus assumed to surround the central engine (if the torus is
roughly perpendicular to the jet). Therefore, only the narrow lines
emitted from low-ionization clouds in the outer regions of the nucleus
are detected in the optical spectrum. Furthermore, the synchrotron
emission from the jet is barely visible and the host galaxy emission
dominates. In the DXRBS an object exhibiting this kind of spectrum is
classified as a Narrow Line Radio Galaxy (NLRG).

A change in the orientation of the jet brings the BLR into sight and
leads, since relativistic effects also take place, to an increase of
the jet synchrotron component. A source exhibiting both broad and
narrow lines is classified in the DXRBS as a quasar. A dividing value
of the full width at half maximum of $1000$ km/s is adopted for quantifying 
the terms ``narrow''
and ``broad.'' The transition in the orientation of the galaxy from
larger to smaller angles is mirrored in a decrease of the value of the
radio spectral index (Orr \& Browne 1982). Based on this, a
distinction between steep spectrum radio quasars (SSRQ) ($\alpha_r \ga
0.5$) and flat spectrum radio quasars (FSRQ) ($\alpha_r \la 0.5$) is
made. A limit of 0.5 is generally chosen, since radio sources become
increasingly core-dominated below this value.  Note that for the
identifications published in Paper I we chose to use a cut at
$\alpha_r =0.70$ to separate FSRQ and SSRQ. In this paper we follow
the commonly accepted division at $\alpha_r =0.50$. However, in the
future we plan to refine this separation by using morphological,
variability, and polarization information.

So far, there is no clear evidence for a stable BLR in FR I
galaxies. Therefore, the face-on version of these objects, which are
classified as BL Lacs, exhibit only weak narrow lines, if any at
all. Broad lines have been observed in the spectra of a few objects of
this class, but only temporarily (e.g., Vermeulen et al. 1995; 
Corbett et al. 1996). 

Therefore, both BL Lac objects and NLRG show typically only narrow
emission lines. To distinguish between these two classes of objects,
we adopt in this publication the March\~a et al. (1996)
definition. These authors base their criterion for an object to be
classified as either a BL Lac or a radio galaxy on the value of the Ca
break, a stellar absorption feature in the optical spectrum defined by
$C = (f_+ - f_-) / f_+$ (where $f_-$ and $f_+$ are the fluxes in the
rest frame wavelength regions $3750 - 3950$ \AA~ and $4050 - 4250$
\AA~ respectively) and the equivalent width (EW) of the strongest
emission line.  In their Fig. 6 March\~a et al. show that objects in a
triangular area limited by Ca break (contrast) $C=0.4$ and a diagonal
line (which assumes the line and continuum emission of the BL Lac
object 3C 371 as its starting point and a smoothly decreasing AGN
contribution) should be called BL Lacs. Candidates with narrow
emission lines and $C > 0.4$, and therefore falling outside this
region, are suggested to be classified as NLRG.

\begin{table*}
\begin{minipage}{160mm}
{\bf Table 4.} Positional Information\\[0.1cm]
\begin{tabular}{lllcllllcc}
\hline
Name & \multicolumn{2}{l}{WGACAT Position} & PSPC & \multicolumn{2}{l}{PMN or GB6 Position} & \multicolumn{2}{l}{Optical Position} & Offset & \multicolumn{1}{c}{Offset/} \\
& RA & DEC & Center & RA & DEC & RA & DEC & (X$-$O) & \multicolumn{1}{c}{Error} \\
& & & Offset & & & & & & \\
& & & [arcmin]& & & & & [arcsec] & \\
(1) & (2) & (3) & (4) & (5) & (6) & (7) & (8) & (9) & (10) \\
\hline
WGAJ0010.5$-$3027 &00 10 33.7 &$-$30 27 14&47.4 &00 10 33.5 &$-$30 27 31&00 10 35.7 &$-$30 27 46&41.1& 1.0 \\
WGAJ0014.5$-$3059 &00 14 34.1 &$-$30 59 22&37.8 &00 14 35.8 &$-$30 59 08&00 14 38.0 &$-$30 59 19&50.2& 1.4 \\
WGAJ0034.4$-$2133 &00 34 29.9 &$-$21 33 24&@9.1 &00 34 30.5 &$-$21 34 03&00 34 30.9 &$-$21 33 53&32.2& 2.3 \\
WGAJ0106.7$-$1034 &01 06 43.1 &$-$10 34 03&36.1 &01 06 43.6 &$-$10 34 26&01 06 44.1 &$-$10 34 10&16.3& 0.4 \\
WGAJ0126.2$-$0500 &01 26 13.6 &$-$05 00 49&28.8 &01 26 09.1 &$-$05 00 40&01 26 15.0 &$-$05 01 23&39.9& 1.4 \\
WGAJ0227.5$-$0847 &02 27 32.1 &$-$08 47 35&37.3 &02 27 32.6 &$-$08 47 59&02 27 32.1 &$-$08 48 13&38.0& 1.0 \\
WGAJ0251.9$-$2051 &02 51 55.2 &$-$20 51 42&15.4 &02 51 55.9 &$-$20 52 18&02 51 54.9 &$-$20 51 49$^a$&@8.2& 0.4 \\
WGAJ0305.3$-$2420 &03 05 20.2 &$-$24 20 39&29.5 &03 05 20.9 &$-$24 21 55&03 05 19.5 &$-$24 21 34&55.8& 1.9 \\
WGAJ0307.7$-$4717 &03 07 42.6 &$-$47 17 20&33.4 &03 07 38.2 &$-$47 17 32&03 07 39.9 &$-$47 17 43&35.8& 1.0 \\
WGAJ0322.1$-$5205 &03 22 08.7 &$-$52 05 32&17.9 &03 22 11.5 &$-$52 06 37&03 22 07.9 &$-$52 05 36&@8.4& 0.5 \\
WGAJ0322.6$-$1335 &03 22 37.8 &$-$13 35 15&12.1 &03 22 39.2 &$-$13 35 33&03 22 38.5 &$-$13 35 18&10.6& 0.6 \\
WGAJ0421.5$+$1433 &04 21 33.4 &$+$14 33 44&11.8 &04 21 33.1 &$+$14 33 43&04 21 33.1 &$+$14 33 54$^b$&10.9& 0.6 \\
WGAJ0427.2$-$0756 &04 27 13.7 &$-$07 56 01&24.2 &04 27 17.9 &$-$07 56 37&04 27 14.2 &$-$07 56 24&24.2& 0.8 \\
WGAJ0438.7$-$4727 &04 38 45.3 &$-$47 27 51&23.1 &04 38 47.4 &$-$47 28 16&04 38 47.0 &$-$47 28 01&19.9& 0.7 \\
WGAJ0513.8$+$0156 &05 13 52.3 &$+$01 56 59&14.9 &05 13 52.7 &$+$01 56 50&05 13 51.8 &$+$01 56 29$^c$&30.6& 1.7 \\
                   &05 13 52.3 &$+$01 56 59&14.9 &05 13 52.7 &$+$01 56 50&05 13 49.4 &$+$01 55 52$^c$&79.9& 1.7 \\
                   &05 13 53.4 &$+$01 56 21&14.3 &05 13 52.7 &$+$01 56 50&05 13 51.8 &$+$01 56 29&25.3& 1.4 \\
                   &05 13 53.4 &$+$01 56 21&14.3 &05 13 52.7 &$+$01 56 50&05 13 49.4 &$+$01 55 52&66.6& 1.4 \\
WGAJ0528.5$-$5820 &05 28 34.7 &$-$58 20 12&33.4 &05 28 33.2 &$-$58 20 25&05 28 34.9 &$-$58 20 20&@8.2& 0.2 \\
WGAJ0533.7$-$5817 &05 33 45.2 &$-$58 17 46&18.9 &05 33 43.3 &$-$58 17 41&05 33 45.3 &$-$58 18 02&16.0& 0.9 \\
WGAJ0534.9$-$6439 &05 34 58.7 &$-$64 39 02&@2.1 &05 35 01.1 &$-$64 38 35&05 35 00.2 &$-$64 38 57&10.8& 0.8 \\
WGAJ0646.8$+$6807 &06 46 49.8 &$+$68 07 46&29.1 &06 46 41.3 &$+$68 07 40&06 46 42.4 &$+$68 07 41$^d$&41.7& 1.4 \\
WGAJ0651.9$+$6955 &06 51 56.3 &$+$69 55 35&34.8 &06 51 55.2 &$+$69 55 16&06 51 54.6 &$+$69 55 26&12.6& 0.3 \\
WGAJ0741.7$-$5304 &07 41 42.2 &$-$53 04 42&29.4 &07 41 49.6 &$-$53 05 16&07 41 49.8 &$-$53 04 39&68.5& 2.4 \\
WGAJ0747.0$-$6744 &07 47 03.6 &$-$67 44 41&@7.1 &07 46 55.7 &$-$67 44 57&07 47 05.0 &$-$67 44 41&@8.0& 0.6 \\
WGAJ0829.5$+$0858 &08 29 30.4 &$+$08 58 25&26.4 &08 29 29.9 &$+$08 58 19&08 29 30.3 &$+$08 58 21&@4.3& 0.1 \\
WGAJ0847.2$+$1133 &08 47 13.0 &$+$11 34 30&48.1 &08 47 10.3 &$+$11 34 24&08 47 12.9 &$+$11 33 50&40.0& 1.0 \\
WGAJ0853.0$+$2004 &08 53 02.3 &$+$20 04 34&23.6 &08 53 03.1 &$+$20 04 22&08 53 02.7 &$+$20 04 22&13.2& 0.5 \\
                   &08 53 02.7 &$+$20 04 23&23.5 &08 53 03.1 &$+$20 04 22&08 53 02.7 &$+$20 04 22&@1.0& 0.0 \\
WGAJ0908.2$+$5031 &09 08 17.4 &$+$50 31 10&19.5 &09 08 16.3 &$+$50 31 06&09 08 16.9 &$+$50 31 05&@6.9& 0.4 \\
WGAJ0927.7$-$0900 &09 27 44.9 &$-$09 00 28&22.7 &09 27 48.9 &$-$09 01 04&09 27 46.9 &$-$09 00 22&30.2& 1.1 \\
WGAJ0931.9$+$5533 &09 31 58.2 &$+$55 33 45&24.1 &09 32 00.8 &$+$55 33 34&09 32 00.2 &$+$55 33 46$^e$&17.0& 0.6 \\
WGAJ0937.1$+$5008 &09 37 10.9 &$+$50 08 54&32.0 &09 37 11.9 &$+$50 08 44&09 37 12.3 &$+$50 08 52&13.6& 0.4 \\
                   &09 37 11.2 &$+$50 09 02&31.9 &09 37 11.9 &$+$50 08 44&09 37 12.3 &$+$50 08 52&14.6& 0.4 \\
                   &09 37 12.1 &$+$50 09 02&24.4 &09 37 11.9 &$+$50 08 44&09 37 12.3 &$+$50 08 52&10.2& 0.4 \\
                   &09 37 12.1 &$+$50 08 53&25.5 &09 37 11.9 &$+$50 08 44&09 37 12.3 &$+$50 08 52&@2.2& 0.1 \\
                   &09 37 13.6 &$+$50 08 56&24.4 &09 37 11.9 &$+$50 08 44&09 37 12.3 &$+$50 08 52&13.1& 0.5 \\
WGAJ1006.1$+$3236 &10 06 08.3 &$+$32 36 22&46.2 &10 06 07.1 &$+$32 36 20&10 06 07.7 &$+$32 36 28&@9.7& 0.2 \\
WGAJ1006.5$+$0509 &10 06 34.1 &$+$05 09 46&35.4 &10 06 37.9 &$+$05 09 50&10 06 37.6 &$+$05 09 54&52.9& 1.5 \\
WGAJ1010.8$-$0201 &10 10 52.6 &$-$02 01 45&40.8 &10 10 52.6 &$-$02 00 57&10 10 51.6 &$-$02 02 27&44.6& 1.1 \\
WGAJ1026.4$+$6746 &10 26 29.1 &$+$67 46 49&40.3 &10 26 33.2 &$+$67 46 02&10 26 34.0 &$+$67 46 12&46.3& 1.1 \\
WGAJ1028.5$-$0236 &10 28 33.2 &$-$02 36 49&38.6 &10 28 34.3 &$-$02 37 12&10 28 34.1 &$-$02 36 60&17.4& 0.5 \\
WGAJ1028.6$-$0336 &10 28 39.4 &$-$03 36 52&35.6 &10 28 39.5 &$-$03 35 57&10 28 40.4 &$-$03 36 19&36.2& 1.0 \\
WGAJ1056.9$-$7649 &10 56 54.3 &$-$76 49 09&51.3 &10 56 57.0 &$-$76 49 23&10 56 54.6 &$-$76 48 35&34.0& 0.6 \\
WGAJ1057.6$-$7724 &10 57 41.0 &$-$77 24 13&26.7 &10 57 31.1 &$-$77 24 24&10 57 33.4 &$-$77 24 29&29.6& 1.0 \\
WGAJ1101.8$+$6241 &11 01 53.2 &$+$62 41 20&39.6 &11 01 53.9 &$+$62 41 55&11 01 53.4 &$+$62 41 50$^f$&30.0& 0.8 \\
WGAJ1105.3$-$1813 &11 05 18.9 &$-$18 13 06&25.0 &11 05 15.2 &$-$18 13 04&11 05 19.2 &$-$18 13 13&@8.2& 0.3 \\
WGAJ1116.1$+$0828 &11 16 08.9 &$+$08 28 14&50.5 &11 16 10.2 &$+$08 29 19&11 16 10.0 &$+$08 29 22&69.9& 1.3 \\
WGAJ1120.4$+$5855 &11 20 26.2 &$+$58 55 57&18.4 &11 20 27.7 &$+$58 56 22&11 20 27.4 &$+$58 56 12&17.6& 1.0 \\
                   &11 20 27.3 &$+$58 56 19&16.3 &11 20 27.7 &$+$58 56 22&11 20 27.4 &$+$58 56 12&@7.0& 0.4 \\
WGAJ1204.2$-$0710 &12 04 16.1 &$-$07 10 03&34.6 &12 04 16.8 &$-$07 10 34&12 04 16.7 &$-$07 10 10&11.3& 0.3 \\
WGAJ1206.2$+$2823 &12 06 17.5 &$+$28 23 19&35.7 &12 06 20.5 &$+$28 23 02&12 06 19.7 &$+$28 22 54&38.3& 1.1 \\
WGAJ1213.0$+$3248 &12 13 03.8 &$+$32 48 00&34.9 &12 13 04.3 &$+$32 47 42&12 13 03.8 &$+$32 47 37&23.0& 0.6 \\
WGAJ1213.2$+$1443 &12 13 14.0 &$+$14 43 47&43.5 &12 13 16.6 &$+$14 44 14&12 13 14.9 &$+$14 44 00&18.4& 0.4 \\
WGAJ1217.1$+$2925 &12 17 09.3 &$+$29 25 28&29.2 &12 17 08.6 &$+$29 25 56&12 17 08.3 &$+$29 25 32&13.7& 0.5 \\
WGAJ1222.6$+$2934 &12 22 39.5 &$+$29 34 40&41.9 &12 22 42.4 &$+$29 34 39&12 22 43.1 &$+$29 34 41&47.0& 1.1 \\
                   &12 22 42.5 &$+$29 34 54&20.1 &12 22 42.4 &$+$29 34 39&12 22 43.1 &$+$29 34 40&16.0& 0.6 \\
WGAJ1223.9$+$0650 &12 23 53.2 &$+$06 50 13&31.2 &12 23 54.7 &$+$06 50 13&12 23 54.7 &$+$06 50 02&24.9& 0.7 \\
\hline
\end{tabular}
\end{minipage}
\end{table*}

\begin{table*}
\begin{minipage}{160mm}
\setcounter{table}{4}
\contcaption{}
\begin{tabular}{lllcllllcc}
\hline
Name & \multicolumn{2}{l}{WGACAT Position} & PSPC & \multicolumn{2}{l}{PMN or GB6 Position} & \multicolumn{2}{l}{Optical Position} & Offset & \multicolumn{1}{c}{Offset/} \\
& RA & DEC & Center & RA & DEC & RA & DEC & (X$-$O) & \multicolumn{1}{c}{Error} \\
& & & Offset & & & & & & \\
& & & [arcmin] & & & & & [arcsec] & \\
(1) & (2) & (3) & (4) & (5) & (6) & (7) & (8) & (9) & (10) \\
\hline
WGAJ1225.5$+$0715 &12 25 30.8 &$+$07 15 59&18.1 &12 25 30.8 &$+$07 15 49&12 25 31.2 &$+$07 15 52&@9.2& 0.5 \\
WGAJ1229.5$+$2711 &12 29 29.0 &$+$27 12 21&40.4 &12 29 33.3 &$+$27 11 57&12 29 34.3 &$+$27 11 57&74.7& 1.8 \\
WGAJ1311.3$-$0521 &13 11 22.3 &$-$05 21 16&30.1 &13 11 21.2 &$-$05 21 05&13 11 17.8 &$-$05 21 22&67.5& 1.9 \\
WGAJ1314.0$-$3304 &13 14 03.0 &$-$33 04 04&12.8 &13 14 04.5 &$-$33 03 07&13 14 03.4 &$-$33 03 56&@9.4& 0.5 \\
WGAJ1315.1$+$2841 &13 15 10.3 &$+$28 41 12&30.1 &13 15 13.9 &$+$28 40 59&13 15 13.6 &$+$28 40 53&47.4& 1.3 \\
WGAJ1320.4$+$0140 &13 20 26.5 &$+$01 40 37&16.1 &13 20 27.1 &$+$01 40 48&13 20 26.8 &$+$01 40 36&@4.6& 0.3 \\
WGAJ1323.8$-$3653 &13 23 47.5 &$-$36 53 06&28.9 &13 23 46.0 &$-$36 52 58&13 23 45.9 &$-$36 53 39&38.2& 1.3 \\
WGAJ1329.0$+$5009 &13 29 00.8 &$+$50 09 39&38.5 &13 29 06.0 &$+$50 09 21&13 29 05.8 &$+$50 09 26&49.8& 1.4 \\
WGAJ1332.7$+$4722 &13 32 46.0 &$+$47 22 32&32.2 &13 32 45.7 &$+$47 22 26&13 32 45.3 &$+$47 22 22&12.3& 0.3 \\
WGAJ1333.1$-$3323 &13 33 07.2 &$-$33 23 52&23.2 &13 33 06.4 &$-$33 24 18&13 33 08.9 &$-$33 24 39&51.6& 1.8 \\
WGAJ1337.2$-$1319 &13 37 13.1 &$-$13 19 03&23.5 &13 37 16.4 &$-$13 19 04&13 37 14.9 &$-$13 19 17&29.8& 1.0 \\
WGAJ1353.2$-$4720 &13 53 14.9 &$-$47 20 56&22.2 &13 53 15.4 &$-$47 20 53&13 53 16.7 &$-$47 20 55&18.3& 0.6 \\
WGAJ1359.6$+$4010 &13 59 36.1 &$+$40 10 54&27.1 &13 59 37.5 &$+$40 11 42&13 59 38.1 &$+$40 11 38&49.6& 1.7 \\
WGAJ1400.7$+$0425 &14 00 46.4 &$+$04 25 51&21.0 &14 00 48.5 &$+$04 25 33&14 00 48.4 &$+$04 25 31&36.0& 1.3 \\
WGAJ1402.7$-$3334 &14 02 43.5 &$-$33 34 01&25.3 &14 02 41.0 &$-$33 34 59&14 02 41.4 &$-$33 34 09&27.4& 0.9 \\
WGAJ1404.2$+$3413 &14 04 15.8 &$+$34 13 21&29.0 &14 04 16.7 &$+$34 13 13&14 04 16.7 &$+$34 13 16&12.2& 0.4 \\
WGAJ1406.9$+$3433 &14 06 54.4 &$+$34 33 42&21.2 &14 06 54.0 &$+$34 33 28&14 06 53.9 &$+$34 33 37&@8.0& 0.3 \\
WGAJ1416.4$+$1242 &14 16 26.8 &$+$12 42 47&40.0 &14 16 28.3 &$+$12 42 26&14 16 28.6 &$+$12 42 13&43.0& 1.0 \\
                   &14 16 28.3 &$+$12 42 18&26.5 &14 16 28.3 &$+$12 42 26&14 16 28.6 &$+$12 42 13&@6.7& 0.2 \\
WGAJ1417.5$+$2645 &14 17 30.2 &$+$26 45 12&@8.6 &14 17 30.6 &$+$26 45 07&14 17 30.4 &$+$26 44 57&15.2& 1.1 \\
WGAJ1419.1$+$0603 &14 19 09.6 &$+$06 03 49&26.6 &14 19 09.2 &$+$06 03 34&14 19 09.3 &$+$06 03 30&19.5& 0.7 \\
WGAJ1420.6$+$0650 &14 20 40.0 &$+$06 50 58&32.1 &14 20 40.9 &$+$06 50 57&14 20 41.0 &$+$06 50 58&14.9& 0.4 \\
                   &14 20 40.6 &$+$06 51 04&32.3 &14 20 40.9 &$+$06 50 57&14 20 41.0 &$+$06 50 58&@6.3& 0.2 \\
WGAJ1423.3$+$4830 &14 23 18.0 &$+$48 30 06&14.1 &14 23 18.4 &$+$48 30 17&14 23 18.0 &$+$48 30 16&10.0& 0.5 \\
WGAJ1427.9$+$3247 &14 27 56.8 &$+$32 47 44&25.9 &14 27 58.3 &$+$32 47 41&14 27 58.8 &$+$32 47 40$^g$&25.5& 0.9 \\
WGAJ1442.3$+$5236 &14 42 21.3 &$+$52 36 17&33.3 &14 42 19.4 &$+$52 36 20&14 42 19.6 &$+$52 36 21&16.0& 0.4 \\
WGAJ1457.7$-$2818 &14 57 42.0 &$-$28 18 43&19.1 &14 57 43.1 &$-$28 19 19&14 57 44.6 &$-$28 19 21&51.2& 2.7 \\
WGAJ1457.9$-$2124 &14 57 55.3 &$-$21 24 56&32.4 &14 57 54.8 &$-$21 25 08&14 57 54.1 &$-$21 24 58&16.9& 0.5 \\
                   &14 57 55.3 &$-$21 24 12&32.3 &14 57 54.8 &$-$21 25 08&14 57 54.1 &$-$21 24 58&49.0& 1.3 \\
WGAJ1506.6$-$4008 &15 06 36.0 &$-$40 08 25&30.0 &15 06 33.7 &$-$40 07 35&15 06 37.2 &$-$40 08 00&28.5& 0.8 \\
WGAJ1507.9$+$6214 &15 07 55.9 &$+$62 14 09&33.5 &15 07 57.9 &$+$62 13 44&15 07 57.3 &$+$62 13 35&35.4& 1.0 \\
WGAJ1509.5$-$4340 &15 09 35.8 &$-$43 40 14&25.8 &15 09 34.1 &$-$43 40 41&15 09 35.7 &$-$43 40 32&18.0& 0.6 \\
WGAJ1539.1$-$0658 &15 39 09.8 &$-$06 58 27&@9.1 &15 39 09.4 &$-$06 58 29&15 39 09.6 &$-$06 58 43&16.3& 1.2 \\
WGAJ1543.6$+$1847 &15 43 40.7 &$+$18 47 41&25.0 &15 43 42.2 &$+$18 47 07&15 43 43.8 &$+$18 47 20&48.8& 1.7 \\
WGAJ1606.0$+$2031 &16 06 05.3 &$+$20 31 43&14.1 &16 06 05.7 &$+$20 32 14&16 06 05.9 &$+$20 32 09&27.3& 1.5 \\
WGAJ1610.3$-$3958 &16 10 20.5 &$-$39 58 22&22.1 &16 10 21.9 &$-$39 58 59&16 10 21.9 &$-$39 58 59&40.3& 1.4 \\
                   &16 10 21.1 &$-$39 58 53&@5.8 &16 10 21.9 &$-$39 58 59&16 10 21.9 &$-$39 58 59&11.0& 0.8 \\
                   &16 10 22.1 &$-$39 58 37&14.0 &16 10 21.9 &$-$39 58 59&16 10 21.9 &$-$39 58 59&22.1& 1.2 \\
                   &16 10 23.9 &$-$39 58 28&23.8 &16 10 21.9 &$-$39 58 59&16 10 21.9 &$-$39 58 59&38.6& 1.3 \\
WGAJ1626.6$+$5809 &16 26 37.3 &$+$58 09 39&32.5 &16 26 36.3 &$+$58 09 14&16 26 37.4 &$+$58 09 16&23.0& 0.6 \\
WGAJ1629.7$+$2117 &16 29 47.0 &$+$21 17 16&14.3 &16 29 47.6 &$+$21 17 22&16 29 47.7 &$+$21 17 16&@9.8& 0.5 \\
WGAJ1648.4$+$4104 &16 48 28.3 &$+$41 04 12&32.1 &16 48 30.0 &$+$41 04 07&16 48 29.3 &$+$41 04 06&12.8& 0.4 \\
WGAJ1656.6$+$5321 &16 56 41.8 &$+$53 21 48&24.3 &16 56 41.3 &$+$53 21 51&16 56 39.7 &$+$53 21 49&18.8& 0.7 \\
WGAJ1656.8$+$6012 &16 56 41.7 &$+$60 12 07&51.4 &16 56 47.8 &$+$60 12 28&16 56 48.3 &$+$60 12 16&50.0& 0.9 \\
                   &16 56 48.1 &$+$60 12 21&42.4 &16 56 47.8 &$+$60 12 28&16 56 48.3 &$+$60 12 16&@5.2& 0.1 \\
WGAJ1722.3$+$3103 &17 22 19.7 &$+$31 03 17&@8.8 &17 22 18.7 &$+$31 03 29&17 22 19.0 &$+$31 03 23&10.8& 0.8 \\
WGAJ1738.6$-$5333 &17 38 37.5 &$-$53 33 54&17.6 &17 38 37.5 &$-$53 34 05&17 38 38.8 &$-$53 34 10&19.7& 1.1 \\
WGAJ1804.7$+$1755 &18 04 42.8 &$+$17 55 34&16.3 &18 04 43.7 &$+$17 55 36&18 04 42.5 &$+$17 55 59&25.4& 1.4 \\
WGAJ1808.2$-$5011 &18 08 13.0 &$-$50 11 38&18.4 &18 08 13.2 &$-$50 11 55&18 08 13.9 &$-$50 11 54&18.2& 1.0 \\
WGAJ1826.1$-$3650 &18 26 07.4 &$-$36 50 37&15.6 &18 26 09.4 &$-$36 51 12&18 26 08.1 &$-$36 50 41&@9.3& 0.5 \\
WGAJ1827.1$-$4533 &18 27 09.9 &$-$45 33 07&23.9 &18 27 11.8 &$-$45 33 02&18 27 10.2 &$-$45 33 00&@4.4& 0.2 \\
WGAJ1834.4$-$5856 &18 34 27.3 &$-$58 56 43&@4.5 &18 34 28.1 &$-$58 56 35&18 34 27.5 &$-$58 56 36&@7.2& 0.5 \\
                   &18 34 27.3 &$-$58 56 11&31.7 &18 34 28.1 &$-$58 56 35&18 34 27.5 &$-$58 56 36&25.0& 0.7 \\
WGAJ1837.7$-$5848 &18 37 44.9 &$-$58 48 05&30.4 &18 37 52.2 &$-$58 48 14&18 37 53.8 &$-$58 48 09&69.2& 1.9 \\
WGAJ1840.9$+$5452 &18 40 58.2 &$+$54 52 29&43.3 &18 40 56.8 &$+$54 52 29&18 40 57.4 &$+$54 52 14&16.5& 0.4 \\
WGAJ1911.8$-$2102 &19 11 50.6 &$-$21 02 21&31.2 &19 11 53.8 &$-$21 02 49&19 11 54.1 &$-$21 02 44&54.5& 1.5 \\
\hline
\end{tabular}
\end{minipage}
\end{table*}

\begin{table*}
\begin{minipage}{160mm}
\setcounter{table}{4}
\contcaption{}
\begin{tabular}{lllcllllcc}
\hline
Name & \multicolumn{2}{l}{WGACAT Position} & PSPC & \multicolumn{2}{l}{PMN or GB6 Position} & \multicolumn{2}{l}{Optical Position} & Offset & \multicolumn{1}{c}{Offset/} \\
& RA & DEC & Center & RA & DEC & RA & DEC & (X$-$O) & \multicolumn{1}{c}{Error} \\
& & & Offset & & & & & & \\
& & & [arcmin] & & & & & [arcsec] & \\
(1) & (2) & (3) & (4) & (5) & (6) & (7) & (8) & (9) & (10) \\
\hline
WGAJ1936.8$-$4719 &19 36 53.1 &$-$47 19 47&42.4 &19 36 55.0 &$-$47 19 38&19 36 56.1 &$-$47 19 50&30.7& 0.7 \\
WGAJ2109.7$-$1332 &21 09 47.0 &$-$13 32 15&21.0 &21 09 51.6 &$-$13 32 59&21 09 49.9 &$-$13 32 46&52.4& 1.8 \\
WGAJ2151.3$-$4233 &21 51 20.6 &$-$42 33 22&37.9 &21 51 23.4 &$-$42 33 41&21 51 21.9 &$-$42 33 34&18.7& 0.5 \\
WGAJ2154.1$-$1501 &21 54 06.2 &$-$15 01 53&47.3 &21 54 08.6 &$-$15 01 54&21 54 07.5 &$-$15 01 31$^h$&27.2& 0.6 \\
                   &21 54 09.9 &$-$15 02 12&47.7 &21 54 08.6 &$-$15 01 54&21 54 07.5 &$-$15 01 31$^h$&54.0& 1.3 \\
WGAJ2159.3$-$1500 &21 59 18.7 &$-$15 00 33&18.8 &21 59 18.3 &$-$15 00 40&21 59 20.2 &$-$15 00 37&22.1& 1.2 \\
                   &21 59 20.1 &$-$15 00 40&19.1 &21 59 18.3 &$-$15 00 40&21 59 20.2 &$-$15 00 37&@3.3& 0.2 \\
WGAJ2201.6$-$5646 &22 01 37.5 &$-$56 46 22&12.8 &22 01 40.4 &$-$56 46 45&22 01 38.4 &$-$56 46 32&12.4& 0.7\\ 
                   &22 01 37.6 &$-$56 46 18&@9.9 &22 01 40.4 &$-$56 46 45&22 01 38.4 &$-$56 46 32&15.5& 1.1 \\
WGAJ2239.7$-$0631 &22 39 46.5 &$-$06 31 42&41.0 &22 39 48.8 &$-$06 31 48&22 39 46.9 &$-$06 31 51$^i$&10.8& 0.3 \\
WGAJ2320.6$+$0032 &23 20 38.5 &$+$00 32 17&31.9 &23 20 39.1 &$+$00 31 16&23 20 38.0 &$+$00 31 39&38.7& 1.1 \\
WGAJ2329.0$+$0834 &23 29 04.0 &$+$08 34 51&22.0 &23 29 05.5 &$+$08 34 06&23 29 05.8 &$+$08 34 15&44.8& 1.6 \\
WGAJ2330.6$-$3724 &23 30 36.9 &$-$37 24 30&35.1 &23 30 36.9 &$-$37 24 43&23 30 35.8 &$-$37 24 36&14.4& 0.4 \\
\hline
\end{tabular}
$^a$ WGAJ0251.9$-$2051: optical counterpart $7^{\prime\prime}$ off NVSS 
position 

$^b$ WGAJ0421.5$+$1433: optical counterpart $9^{\prime\prime}$ off NVSS 
position; NVSS source extended 

$^c$ WGAJ0513.8$+$0156: three NVSS sources; the northermost optical
counterpart is in between the two strongest sources, while the southernmost
counterpart is $5^{\prime\prime}$ off the faintest NVSS source. See \S~3.4

$^d$ WGAJ0646.8$+$6807: optical counterpart $5^{\prime\prime}$ off NVSS 
position 

$^e$ WGAJ0931.9$+$5533: three NVSS sources; optical counterpart in between
two of them; possible double-lobed source 

$^f$ WGAJ1101.8$+$6241: optical counterpart $6^{\prime\prime}$ off NVSS 

$^g$ WGAJ1427.9$+$3247: two NVSS sources; optical counterpart in between
them; possible double-lobed source 

$^h$ WGAJ2154.1$-$1501: ATCA position; NVSS position $15^{\prime\prime}$ off

$^i$ WGAJ2239.7$-$0631: optical counterpart $7^{\prime\prime}$ off NVSS 
position

\end{minipage}
\end{table*}

\begin{table*}
\begin{minipage}{180mm}
{\bf Table 5.} Object Properties\\[0.1cm]
\begin{tabular}{rcrccccllllll}
\hline
\multicolumn{1}{l}{Name} & \multicolumn{2}{c}{ROSAT} & F$_{0.1-2.0 keV}$ & F$_{1keV}$ & log & F$_{6cm}$ & \multicolumn{1}{c}{$\alpha_{\rm r}$} & \multicolumn{1}{c}{B$_j$} & \multicolumn{1}{c}{O} & \multicolumn{1}{c}{E} & Class & \multicolumn{1}{c}{z}\\
& cts & \multicolumn{1}{c}{hr} & [${\rm erg/cm^{2}/s}$] & [$\mu$Jy] & F$_{\rm X}$/F$_{\rm R}$ & [mJy] &&&&&& \\
&&&&&&&&&&&& \\
\multicolumn{1}{l}{(1)} & (2) & \multicolumn{1}{c}{(3)} & (4) & (5) & (6) & (7) & \multicolumn{1}{c}{(8)} & \multicolumn{1}{c}{(9)} & \multicolumn{1}{c}{(10)} & \multicolumn{1}{c}{(11)} & (12) & \multicolumn{1}{c}{(13)} \\
\hline
WGAJ0010.5$-$3027 &0.022&0.68&0.14E-12&0.011&$-$12.72&391&$-$0.15&$\g$19.1&$\g$19.5&$\g$18.8&FSRQ&1.19@$\pm$0.01  \\     
WGAJ0014.5$-$3059 &0.015&1.00&0.13E-12&0.022&$-$12.08&140&$\p$0.25&$\g$ ...&$\g$19.7&$\g$18.9&FSRQ&2.785$\pm$0.002 \\     
WGAJ0034.4$-$2133 &0.003&2.60&0.26E-13&0.009&$-$12.17&@82&$\p$0.56$^a$&$\g$ ...&$\g$ ...&$\g$22.9$^b$&SSRQ&0.764$\pm$0.001 \\   
WGAJ0106.7$-$1034 &0.039&0.90&0.44E-12&0.091&$-$11.63&188&$\p$0.35&$\g$17.0&$\g$17.7&$\g$17.7&FSRQ&0.469$\pm$0.001 \\     
WGAJ0126.2$-$0500 &0.027&1.25&0.31E-12&0.076&$-$11.19&@54&$\p$0.21&$\g$18.3&$\g$ ...&$\g$ ...&FSRQ&0.411$\pm$0.001 \\     
WGAJ0227.5$-$0847 &0.019&1.07&0.22E-12&0.051&$-$11.69&115&$-$0.34$^c$&$\g$17.3&$\g$ ...&$\g$ ...&FSRQ&2.228$\pm$0.001 \\     
WGAJ0251.9$-$2051 &0.015&1.67&0.13E-12&0.037&$-$11.45&@53&$\p$0.65$^d$&$\g$18.5&$\g$ ...&$\g$ ...&SSRQ&0.761 \\               
WGAJ0305.3$-$2420 &0.013&0.67&0.10E-12&0.008&$-$12.25&102&$-$0.15&$\g$18.8&$\g$17.9&$\g$16.7&NLRG?&0.211$\pm$0.001 \\   
WGAJ0307.7$-$4717 &0.024&0.93&0.22E-12&0.042&$-$11.76&123&$\p$0.70&$\g$23.2&$\g$ ...&$\g$ ...&SSRQ&0.599$\pm$0.001 \\     
WGAJ0322.1$-$5205 &0.156&1.08&0.12E-11&0.262&$-$10.48&@40&$-$0.27&$\g$16.8&$\g$ ...&$\g$ ...&FSRQ&0.416$\pm$0.001 \\     
WGAJ0322.6$-$1335 &0.004&0.64&0.47E-13&0.008&$-$12.58&164&$\p$0.38&$\g$ ...&$\g$22.0$^e$&$\g$ ...&FSRQ&1.468$\pm$0.001 \\     
WGAJ0421.5$+$1433 &0.003&1.85&0.35E-13&0.017&$-$12.42&114&$\p$0.63&$\g$ ...&$\g$18.8&$\g$15.2&NLRG?&0.059$\pm$0.001 \\     
WGAJ0427.2$-$0756 &0.012&1.29&0.14E-12&0.042&$-$11.52&@56&$\p$0.27&$\g$21.0&$\g$ ...&$\g$ ...&FSRQ&1.375$\pm$0.003 \\     
WGAJ0438.7$-$4727 &0.023&1.18&0.20E-12&0.041&$-$11.78&130&$\p$0.54&$\g$20.5&$\g$20.5&$\g$19.7&SSRQ&1.445$\pm$0.004 \\     
WGAJ0513.8$+$0156 &0.031&3.47&0.42E-12&0.123&$-$11.26&131&$\p$0.45&$\g$ ...&$\g$ ...&$\g$14.3&NLRG&0.092$\pm$0.001 \\     
                  &0.023&2.37&0.30E-12&0.095&$-$11.49&&&&&&& \\
                  &0.031&3.47&0.42E-12&0.123&$-$11.26&131&$\p$0.45&$\g$ ...&$\g$ ...&$\g$17.7&NLRG&0.087$\pm$0.001 \\     
                  &0.023&2.37&0.30E-12&0.095&$-$11.49&&&&&&& \\     
WGAJ0528.5$-$5820 &0.014&1.09&0.14E-12&0.031&$-$11.80&@99&$\p$0.70&$\g$18.7&$\g$17.8&$\g$16.7&BL Lac&0.254$\pm$0.001 \\   
$\ast$WGAJ0533.7$-$5817 &0.009&1.70&0.11E-12&0.029&$-$11.68&@66&$\p$0.94&$\g$17.9&$\g$18.5&$\g$18.4&SSRQ&0.757$\pm$0.001 \\     
$\ast$WGAJ0534.9$-$6439 &0.016&1.63&0.21E-12&0.057&$-$11.48&@76&$\p$0.77&$\g$20.5&$\g$ ...&$\g$ ...&SSRQ&1.463$\pm$0.001 \\     
WGAJ0646.8$+$6807 &0.008&0.42&0.89E-13&0.009&$-$12.03&@72&$\p$0.56&$\g$ ...&$\g$ ...&$\g$19.7$^f$&SSRQ&0.927$\pm$0.001 \\     
WGAJ0651.9$+$6955 &0.017&1.39&0.21E-12&0.067&$-$11.67&132&$\p$0.63&$\g$ ...&$\g$20.0&$\g$19.3&SSRQ&1.367 \\               
$\ast$WGAJ0741.7$-$5304 &0.005&1.08&0.71E-13&0.024&$-$11.79&@44&$\p$0.77&$\g$22.2&$\g$ ...&$\g$ ...&SSRQ&3.743$\pm$0.001 \\     
WGAJ0747.0$-$6744 &0.005&1.25&0.68E-13&0.023&$-$11.59&@27&$\p$0.16&$\g$19.7&$\g$ ...&$\g$ ...&FSRQ&1.025 \\               
WGAJ0829.5$+$0858 &0.019&1.24&0.24E-12&0.058&$-$11.82&180&$\p$0.60&$\g$ ...&$\g$21.5$^g$&$\g$ ...&SSRQ&0.866$\pm$0.001 \\     
$\ast$WGAJ0847.2$+$1133 &1.290&1.20&0.16E-10&3.954&$-$@9.24&@32&$\p$0.03&$\g$ ...&$\g$17.8&$\g$16.6&BL Lac&0.199$\pm$0.001 \\     
WGAJ0853.0$+$2004 &0.010&1.20&0.86E-13&0.022&$-$11.73&@60&$\p$0.60&$\g$ ...&$\g$18.2&$\g$18.0&SSRQ&1.930$\pm$0.003 \\     
                  &0.010&1.11&0.83E-13&0.019&$-$11.80&&&&&&& \\     
WGAJ0908.2$+$5031 &0.008&1.00&0.63E-13&0.010&$-$12.20&@86&$\p$0.59&$\g$ ...&$\g$20.8&$\g$19.6&SSRQ&0.917 \\               
WGAJ0927.7$-$0900 &0.014&1.86&0.20E-12&0.052&$-$11.82&181&$-$0.14&$\g$21.3&$\g$ ...&$\g$ ...&FSRQ&0.254$\pm$0.001 \\     
WGAJ0931.9$+$5533 &0.045&1.25&0.43E-12&0.105&$-$11.00&@53&$\p$0.66&$\g$ ...&$\g$16.7&$\g$16.0&SSRQ&0.266$\pm$0.001 \\     
$\ast$WGAJ0937.1$+$5008 &0.056&1.00&0.51E-12&0.063&$-$11.95&315&$-$0.42&$\g$ ...&$\g$19.6&$\g$18.0&FSRQ&0.275$\pm$0.001 \\     
                   &0.069&1.64&0.74E-12&0.167&$-$11.52&&&&&&& \\     
                   &0.056&1.73&0.53E-12&0.127&$-$11.64&&&&&&& \\     
                   &0.072&1.67&0.67E-12&0.160&$-$11.53&&&&&&& \\     
                   &0.058&1.20&0.50E-12&0.082&$-$11.85&&&&&&& \\     
WGAJ1006.1$+$3236 &0.052&0.71&0.43E-12&0.046&$-$11.91&231&$\p$0.62&$\g$ ...&$\g$18.7&$\g$18.1&SSRQ&1.02@$\pm$0.01  \\     
WGAJ1006.5$+$0509 &0.015&1.22&0.16E-12&0.037&$-$11.99&179&$\p$0.12&$\g$ ...&$\g$21.7&$\g$ ...&FSRQ&1.216$\pm$0.006 \\     
WGAJ1010.8$-$0201 &0.063&0.90&0.82E-12&0.164&$-$12.02&826&$\p$0.32&$\g$18.9&$\g$19.9&$\g$18.9&FSRQ&0.896$\pm$0.002 \\     
WGAJ1026.4$+$6746 &0.034&0.90&0.38E-12&0.069&$-$11.55&129&$\p$0.49&$\g$ ...&$\g$16.9$^h$&$\g$17.9$^h$&FSRQ&1.181 \\               
WGAJ1028.5$-$0236 &0.019&1.16&0.24E-12&0.068&$-$11.53&@94&$-$0.06&$\g$21.9&$>$22.0&$\g$19.6&FSRQ&0.476$\pm$0.002 \\     
$\ast$WGAJ1028.6$-$0336 &0.008&0.58&0.76E-13&0.009&$-$11.99&@61&$\p$0.74&$\g$19.1&$\g$20.5&$\g$ ...&SSRQ&1.781$\pm$0.001 \\     
WGAJ1056.9$-$7649 &0.019&1.21&0.34E-12&0.113&$-$11.38&@98&$\p$0.00&$\g$20.9&$\g$ ...&$\g$ ...&NLRG?&0.193$\pm$0.001 \\     
$\ast$WGAJ1057.6$-$7724 &0.005&0.95&0.71E-13&0.020&$-$12.76&431&$\p$0.71&$\g$ ...&$\g$21.4$^i$&$\g$ ...&NLRG?&0.181$\pm$0.001 \\     
WGAJ1101.8$+$6241 &0.070&0.82&0.46E-12&0.038&$-$12.46&693&$\p$0.12&$\g$ ...&$\g$17.9&$\g$17.7&FSRQ&0.663$\pm$0.001 \\     
WGAJ1105.3$-$1813 &0.006&1.50&0.80E-13&0.023&$-$11.77&@59&$-$0.77&$\g$ ...&$\g$18.5&$\g$20.1&FSRQ&0.578$\pm$0.001 \\     
WGAJ1116.1$+$0828 &0.014&1.26&0.22E-12&0.063&$-$11.98&282&$-$0.08&$\g$ ...&$\g$22.4$^j$&$\g$ ...&FSRQ&0.486  \\     
$\ast$WGAJ1120.4$+$5855 &0.008&0.96&0.60E-13&0.005&$-$12.19&@46&$\p$0.61&$\g$ ...&$\g$18.5&$\g$16.0&NLRG&0.158$\pm$0.001 \\     
                   &0.026&1.46&0.21E-12&0.033&$-$11.40&&&&&&& \\     
WGAJ1204.2$-$0710 &0.058&0.76&0.45E-12&0.070&$-$11.50&128&$\p$0.22&$\g$18.0&$\g$ ...&$\g$ ...&BL Lac&0.185$\pm$0.001 \\   
$\ast$WGAJ1206.2$+$2823 &0.013&0.31&0.89E-13&0.002&$-$11.96&@21&$\p$0.41&$\g$ ...&$\g$19.5&$\g$18.9&FSRQ&0.708$\pm$0.002 \\     
WGAJ1213.0$+$3248 &0.015&1.15&0.12E-12&0.022&$-$11.70&@61&$\p$0.70&$\g$ ...&$\g$19.9&$\g$19.2&SSRQ&2.502$\pm$0.003 \\ 
WGAJ1213.2$+$1443 &0.044&1.06&0.48E-12&0.100&$-$11.08&@61&$\p$0.50&$\g$ ...&$\g$21.0&$\g$19.9&FSRQ&0.718$\pm$0.002 \\   
$\ast$WGAJ1217.1$+$2925 &0.012&0.92&0.94E-13&0.014&$-$11.80&@49&$\p$0.27&$\g$ ...&$\g$19.9&$\g$19.2&FSRQ&0.974 \\     
WGAJ1222.6$+$2934 &0.016&1.02&0.15E-12&0.029&$-$11.59&@60&$\p$0.55&$\g$ ...&$\g$19.5&$\g$18.7&SSRQ&0.787$\pm$0.001 \\     
                  &0.016&1.33&0.15E-12&0.033&$-$11.54&&&&&&& \\     
WGAJ1223.9$+$0650 &0.018&1.01&0.18E-12&0.030&$-$12.18&251&$\p$0.11&$\g$ ...&$\g$21.2&$\g$ ...&FSRQ&1.189$\pm$0.003 \\  
WGAJ1225.5$+$0715 &0.010&1.32&0.84E-13&0.016&$-$11.96&@75&$\p$0.54&$\g$ ...&$\g$21.3&$>$20.0&SSRQ&1.12@$\pm$0.01  \\    
\hline
\end{tabular}
\end{minipage}
\end{table*}

\begin{table*}
\setcounter{table}{5}
\begin{minipage}{180mm}
\contcaption{}
\begin{tabular}{rcrccccllllll}
\hline
\multicolumn{1}{l}{Name} & \multicolumn{2}{c}{ROSAT} & F$_{0.1-2.0 keV}$ & F$_{1keV}$ & log & F$_{6cm}$ & \multicolumn{1}{c}{$\alpha_{\rm r}$} & \multicolumn{1}{c}{B$_j$} & \multicolumn{1}{c}{O} & \multicolumn{1}{c}{E} & Class & \multicolumn{1}{c}{z}\\
& cts & \multicolumn{1}{c}{hr} & [${\rm erg/cm^{2}/s}$] & [$\mu$Jy] & F$_{\rm X}$/F$_{\rm R}$ & [mJy] &&&&&& \\
&&&&&&&&&&&& \\
\multicolumn{1}{l}{(1)} & (2) & \multicolumn{1}{c}{(3)} & (4) & (5) & (6) & (7) & \multicolumn{1}{c}{(8)} & \multicolumn{1}{c}{(9)} & \multicolumn{1}{c}{(10)} & \multicolumn{1}{c}{(11)} & (12) & \multicolumn{1}{c}{(13)} \\
\hline
WGAJ1229.5$+$2711 &0.024&1.67&0.28E-12&0.079&$-$11.56&165&$\p$0.16&$\g$ ...&$>$22.0&$\g$19.2&NLRG&0.490$\pm$0.001 \\
WGAJ1311.3$-$0521 &0.027&0.80&0.28E-12&0.040&$-$11.31&@46&$\p$0.36&$\g$18.7&$\g$ ...&$\g$ ...&BL Lac&0.160$\pm$0.001 \\
WGAJ1314.0$-$3304 &0.026&1.64&0.33E-12&0.092&$-$11.50&125&$\p$0.50$^k$&$\g$19.3&$\g$ ...&$\g$ ...&FSRQ&0.484$\pm$0.001 \\   
WGAJ1315.1$+$2841 &0.017&0.63&0.12E-12&0.008&$-$12.20&@95&$\p$0.36&$\g$ ...&$\g$20.2&$\g$19.9&FSRQ&1.576$\pm$0.001 \\   
WGAJ1320.4$+$0140 &0.022&1.41&0.20E-12&0.036&$-$12.46&541&$\p$0.01&$\g$20.8&$\g$20.8&$>$20.0&BL Lac&1.235$\pm$0.004 \\             
WGAJ1323.8$-$3653 &0.007&0.76&0.92E-13&0.019&$-$12.18&135&$-$0.04&$\g$21.9&$\g$ ...&$\g$ ...&BL Lac? & ? \\              
WGAJ1329.0$+$5009 &0.023&0.84&0.16E-12&0.018&$-$12.08&133&$\p$0.68&$\g$ ...&$\g$20.1&$\g$19.2&SSRQ&2.65@$\pm$0.01  \\   
WGAJ1332.7$+$4722 &0.033&1.06&0.33E-12&0.056&$-$12.05&333&$\p$0.54&$\g$ ...&$\g$19.3&$\g$18.9&SSRQ&0.668$\pm$0.001 \\   
WGAJ1333.1$-$3323 &0.005&1.22&0.55E-13&0.014&$-$12.33&140&$\p$0.22&$\g$21.7&$\g$ ...&$\g$ ...&FSRQ&2.24@$\pm$0.01  \\   
WGAJ1337.2$-$1319 &0.022&1.22&0.32E-12&0.085&$-$11.30&@71&$\p$0.38$^l$&$\g$ ...&$\g$19.8&$\g$18.8&FSRQ&3.475$\pm$0.003 \\   
WGAJ1353.2$-$4720 &0.019&2.78&0.30E-12&0.128&$-$11.63&198&$\p$0.66&$\g$21.4&$\g$ ...&$\g$ ...&SSRQ&0.550$\pm$0.001 \\   
WGAJ1359.6$+$4010 &0.014&1.20&0.13E-12&0.020&$-$12.41&281&$-$0.76&$\g$ ...&$\g$19.2&$\g$17.6&FSRQ&0.407$\pm$0.001 \\   
WGAJ1400.7$+$0425 &0.009&0.63&0.80E-13&0.008&$-$12.73&267&$\p$0.06&$\g$ ...&$\g$21.3&$\g$ ...&FSRQ&2.55@$\pm$0.01  \\   
WGAJ1402.7$-$3334 &0.042&2.12&0.61E-12&0.172&$-$11.00&@88&$-$0.02&$\g$23.0&$\g$ ...&$\g$ ...&FSRQ&2.14@$\pm$0.01  \\   
WGAJ1404.2$+$3413 &0.011&1.07&0.78E-13&0.012&$-$11.98&@62&$\p$0.67&$\g$ ...&$\g$18.6&$\g$17.6&SSRQ&0.937$\pm$0.002 \\   
WGAJ1406.9$+$3433 &0.014&0.82&0.11E-12&0.010&$-$12.51&204&$\p$0.27&$\g$ ...&$\g$18.7&$\g$17.9&FSRQ&2.556$\pm$0.001 \\   
WGAJ1416.4$+$1242 &0.100&0.56&0.95E-12&0.058&$-$11.36&@98&$\p$0.30&$\g$ ...&$\g$18.1&$\g$17.7&FSRQ&0.335$\pm$0.001 \\   
                  &0.098&1.07&0.87E-12&0.148&$-$11.09&&&&&&& \\   
WGAJ1417.5$+$2645 &0.006&1.47&0.51E-13&0.010&$-$12.17&@77&$\p$0.29&$\g$ ...&$\g$21.9&$>$20.0&FSRQ&1.455$\pm$0.005 \\   
WGAJ1419.1$+$0603 &0.009&1.25&0.97E-13&0.022&$-$12.33&240&$\p$0.01&$\g$ ...&$\g$ ...&$\g$20.5$^m$&FSRQ&2.389$\pm$0.003 \\   
WGAJ1420.6$+$0650 &0.028&1.26&0.33E-12&0.076&$-$11.80&241&$\p$0.56&$\g$ ...&$\g$ ...&$\g$17.5&SSRQ&0.236$\pm$0.001 \\   
                  &0.033&0.73&0.35E-12&0.042&$-$12.00&&&&&&&\\   
WGAJ1423.3$+$4830 &0.022&1.46&0.20E-12&0.037&$-$11.72&100&$\p$0.55&$\g$ ...&$\g$19.4&$\g$19.1&SSRQ&0.569$\pm$0.001 \\   
WGAJ1427.9$+$3247 &0.052&0.99&0.38E-12&0.055&$-$11.32&@65&$\p$0.66&$\g$ ...&$\g$18.1&$\g$17.6&SSRQ&0.568$\pm$0.001 \\   
WGAJ1442.3$+$5236 &0.016&1.35&0.14E-12&0.036&$-$11.77&111&$\p$0.32&$\g$ ...&$\g$19.3&$\g$18.5&FSRQ&1.80@$\pm$0.01  \\   
$\ast$WGAJ1457.7$-$2818 &0.007&0.67&0.89E-13&0.019&$-$12.19&136&$\p$0.40$^n$&$\g$18.8&$\g$ ...&$\g$ ...&FSRQ&1.999$\pm$0.001 \\   
WGAJ1457.9$-$2124 &0.035&1.19&0.54E-12&0.166&$-$11.38&146&$\p$0.79&$\g$21.3&$\g$21.6&$\g$18.3&NLRG&0.319$\pm$0.001 \\   
                  &0.040&1.21&0.61E-12&0.191&$-$11.32&&&&&&& \\   
WGAJ1506.6$-$4008 &0.018&0.68&0.27E-12&0.064&$-$11.60&110&$\p$0.10&$\g$ ...&$\g$19.6$^o$&$\g$ ...&FSRQ&1.031$\pm$0.002 \\   
WGAJ1507.9$+$6214 &0.018&0.56&0.12E-12&0.008&$-$12.52&213&$\p$0.69&$\g$ ...&$\g$18.5&$\g$18.1&SSRQ&1.478$\pm$0.001 \\   
WGAJ1509.5$-$4340 &0.016&1.07&0.24E-12&0.081&$-$11.83&165&$-$0.40&$\g$19.2&$\g$ ...&$\g$ ...&FSRQ&0.776$\pm$0.001 \\   
WGAJ1539.1$-$0658 &0.005&0.94&0.70E-13&0.018&$-$12.03&@76&$-$0.27&$\g$20.0&$\g$ ...&$\g$ ...&BL Lac?& ? \\              
WGAJ1543.6$+$1847 &0.011&1.33&0.13E-12&0.035&$-$12.26&300&$\p$0.13&$\g$ ...&$\g$19.5&$\g$18.8&FSRQ&1.396$\pm$0.006 \\   
WGAJ1606.0$+$2031 &0.006&2.17&0.66E-13&0.018&$-$12.31&196&$\p$0.32&$\g$ ...&$\g$ ...&$\g$18.3&FSRQ&0.383$\pm$0.001 \\   
$\ast$WGAJ1610.3$-$3958 &0.026&3.75&0.40E-12&0.209&$-$12.09&882&$\p$0.19&$\g$20.9&$\g$ ...&$\g$ ...&FSRQ&0.518$\pm$0.001 \\   
                  &0.011&12.00&0.17E-12&0.086&$-$12.69&&&&&&&  \\   
                  &0.023&3.25&0.29E-12&0.165&$-$12.31&&&&&&&  \\   
                  &0.022&2.14&0.30E-12&0.208&$-$12.34&&&&&&&  \\   
WGAJ1626.6$+$5809 &0.032&1.22&0.32E-12&0.067&$-$11.95&315&$\p$0.55&$\g$ ...&$\g$ ...&$\g$17.5&SSRQ&0.748$\pm$0.001 \\   
WGAJ1629.7$+$2117 &0.013&1.38&0.14E-12&0.035&$-$11.81&105&$\p$0.50&$\g$ ...&$\g$21.5&$>$20.0&FSRQ&0.833$\pm$0.002 \\   
WGAJ1648.4$+$4104 &0.061&1.11&0.61E-12&0.109&$-$11.54&197&$\p$0.52&$\g$ ...&$\g$19.8&$\g$19.0&SSRQ&0.851$\pm$0.001 \\   
WGAJ1656.6$+$5321 &0.014&1.27&0.15E-12&0.033&$-$11.94&145&$\p$0.09&$\g$ ...&$\g$ ...&$\g$19.1$^p$&FSRQ&1.555 \\             
WGAJ1656.8$+$6012 &0.032&0.66&0.29E-12&0.029&$-$11.99&184&$-$0.33&$\g$ ...&$\g$19.0$^q$&$\g$18.4$^r$&FSRQ&0.623$\pm$0.001 \\   
                  &0.032&0.38&0.23E-12&0.007&$-$12.37&&&&&&&  \\   
WGAJ1722.3$+$3103 &0.014&1.38&0.16E-12&0.036&$-$11.50&@53&$\p$0.58&$\g$ ...&$\g$21.2&$\g$18.7&SSRQ&0.305$\pm$0.001 \\   
WGAJ1738.6$-$5333 &0.015&1.80&0.19E-12&0.072&$-$12.10&307&$-$0.10&$\g$ ...&$\g$18.4$^r$&$\g$ ...&FSRQ&1.721$\pm$0.001 \\   
WGAJ1804.7$+$1755 &0.009&1.56&0.11E-12&0.036&$-$11.84&@92&$\p$0.29&$\g$ ...&$\g$19.4&$\g$18.7&FSRQ&0.435$\pm$0.001 \\   
WGAJ1808.2$-$5011 &0.011&1.27&0.14E-12&0.042&$-$12.44&425&$-$0.33&$\g$ ...&$\g$20.4$^s$&$\g$ ...&FSRQ&1.606$\pm$0.004 \\   
WGAJ1826.1$-$3650 &0.016&1.72&0.20E-12&0.078&$-$12.35&552&$-$0.06&$\g$ ...&$\g$ ...&$\g$18.8$^t$&FSRQ&0.888 \\             
WGAJ1827.1$-$4533 &0.023&1.16&0.33E-12&0.104&$-$12.09&451&$\p$0.00&$\g$18.1&$\g$ ...&$\g$ ...&FSRQ&1.244$\pm$0.001 \\   
WGAJ1834.4$-$5856 &0.011&2.59&0.15E-12&0.046&$-$12.25&399&$\p$0.00&$\g$19.7&$\g$ ...&$\g$ ...&BL Lac& ? \\ 
                  &0.023&0.95&0.35E-12&0.102&$-$12.03&&&&&&&  \\ 
WGAJ1837.7$-$5848 &0.012&0.80&0.19E-12&0.053&$-$12.02&211&$-$0.08&$\g$19.3&$\g$ ...&$\g$ ...&FSRQ&3.040$\pm$0.002 \\   
WGAJ1840.9$+$5452 &0.040&1.33&0.58E-12&0.179&$-$11.53&252&$\p$0.39&$\g$ ...&$\g$19.4&$\g$18.8&BL Lac&0.646$\pm$0.001 \\ 
WGAJ1911.8$-$2102 &0.018&0.59&0.30E-12&0.084&$-$12.17&443&$\p$0.28&$\g$ ...&$\g$17.0$^u$&$\g$19.5$^u$&FSRQ&1.42@$\pm$0.01  \\   
WGAJ1936.8$-$4719 &0.202&1.36&0.34E-11&0.936&$-$10.17&@60&$-$0.11&$\g$20.3&$\g$ ...&$\g$ ...&BL Lac&0.265$\pm$0.001 \\ 
WGAJ2109.7$-$1332 &0.014&1.01&0.16E-12&0.034&$-$11.49&@51&$\p$0.16&$\g$19.3&$\g$18.2&$\g$17.8&FSRQ&1.226$\pm$0.001 \\
\hline
\end{tabular}
\end{minipage}
\end{table*}

\subsection{Optical Spectra}    

In the Appendix, we present the spectra of the optical
counterparts. All spectra have been smoothed with Gaussian curves of
width 3 pixels.

\begin{table*}
\setcounter{table}{5}
\begin{minipage}{180mm}
\contcaption{}
\begin{tabular}{rcrccccllllll}
\hline
\multicolumn{1}{l}{Name} & \multicolumn{2}{c}{ROSAT} & F$_{0.1-2.0 keV}$ & F$_{1keV}$ & log & F$_{6cm}$ & \multicolumn{1}{c}{$\alpha_{\rm r}$} & \multicolumn{1}{c}{B$_j$} & \multicolumn{1}{c}{O} & \multicolumn{1}{c}{E} & Class & \multicolumn{1}{c}{z}\\
& cts & \multicolumn{1}{c}{hr} & [${\rm erg/cm^{2}/s}$] & [$\mu$Jy] & F$_{\rm X}$/F$_{\rm R}$ & [mJy] &&&&&& \\
&&&&&&&&&&&& \\
\multicolumn{1}{l}{(1)} & (2) & \multicolumn{1}{c}{(3)} & (4) & (5) & (6) & (7) & \multicolumn{1}{c}{(8)} & \multicolumn{1}{c}{(9)} & \multicolumn{1}{c}{(10)} & \multicolumn{1}{c}{(11)} & (12) & \multicolumn{1}{c}{(13)} \\
\hline
WGAJ2151.3$-$4233 &0.030&0.83&0.26E-12&0.031&$-$12.16&265&$\p$0.51$^v$&$\g$16.1$^w$&$\g$ ...&$\g$ ...&NLRG&0.061$\pm$0.001 \\        
WGAJ2154.1$-$1501 &0.033&0.84&0.36E-12&0.072&$-$11.78&219&$\p$0.30&$\g$ ...&$\g$16.5&$\g$16.4&FSRQ&1.208$\pm$0.005 \\   
                  &0.033&1.06&0.35E-12&0.091&$-$11.71&&&&&&& \\
WGAJ2159.3$-$1500 &0.005&1.46&0.49E-13&0.013&$-$12.16&@86&$\p$0.01$^x$&$\g$18.2&$\g$18.9&$\g$18.4&FSRQ&2.270$\pm$0.010 \\   
                   &0.013&1.05&0.13E-12&0.027&$-$11.82&&&&&&&\\
$\ast$WGAJ2201.6$-$5646 &0.047&1.09&0.45E-12&0.077&$-$11.23&@67&$\p$0.71&$\g$17.5&$\g$ ...&$\g$ ...&SSRQ&0.410$\pm$0.001 \\               
                   &0.022&1.47&0.22E-12&0.049&$-$11.44&&&&&&& \\               
WGAJ2239.7$-$0631 &0.069&1.31&0.97E-12&0.259&$-$10.72&@64&$\p$0.54&$\g$ ...&$\g$20.1&$\g$18.8&SSRQ&0.264$\pm$0.001 \\     
WGAJ2320.6$+$0032 &0.013&1.05&0.16E-12&0.036&$-$11.71&@87&$\p$0.45&$\g$18.9&$\g$ ...&$\g$ ...&FSRQ&1.894$\pm$0.001 \\     
WGAJ2329.0$+$0834 &0.004&2.67&0.61E-13&0.020&$-$12.35&273&$\p$0.06&$\g$ ...&$\g$20.5$^y$&$\g$ ...&FSRQ&0.948$\pm$0.001 \\     
WGAJ2330.6$-$3724 &0.032&0.88&0.27E-12&0.038&$-$12.05&238&$\p$0.19&$\g$16.6&$\g$ ...&$\g$ ...&BL Lac&0.279$\pm$0.001\\   
\hline
\end{tabular}
$\ast$ Source does not meet strict DXRBS selection criteria 

$^a$ WGAJ0034.4$-$2133: $\alpha_{\rm ATCA} = 1.91$

$^b$ WGAJ0034.4$-$2133: $V_{\rm mag}$ from ESO image 
	
$^c$ WGAJ0227.5$-$0847: $\alpha_{\rm ATCA} = 0.36$
	
$^d$ WGAJ0251.9$-$2051: $\alpha_{\rm ATCA} = 1.70$
	
$^e$ WGAJ0322.6$-$1335: $V_{\rm mag}$ from ESO image 
	
$^f$ WGAJ0646.8$+$6807: $R_{\rm mag}$ from WYIN image 
	
$^g$ WGAJ0829.5$+$0858: $V_{\rm mag}$ from ESO image 
	
$^h$ WGAJ1026.4$+$6746: two merged objects in APM: given magnitudes assume
a flux ratio of $\sim 2$  

$^i$ WGAJ1057.6$-$7724: $V_{\rm mag}$ from ESO image 
	
$^j$ WGAJ1116.1$+$0828: $V_{\rm mag}$ from ESO image 
	
$^k$ WGAJ1314.0$-$3304: $\alpha_{\rm ATCA} = 0.97$
	
$^l$ WGAJ1337.2$-$1319: $\alpha_{\rm ATCA} = 0.87$ 
	
$^m$ WGAJ1419.1$+$0603: $R_{\rm mag}$ from CTIO image
	
$^n$ WGAJ1457.7$-$2818: $\alpha_{\rm ATCA} = 0.72$, TEXAS survey double 
	
$^o$ WGAJ1506.6$-$4008: $V_{\rm mag}$ from ESO image 
	
$^p$ WGAJ1656.6$+$5321: $R_{\rm mag}$ from WYIN image  
	
$^q$ WGAJ1656.8$+$6012: USNO magnitudes (two merged objects in APM)
	
$^r$ WGAJ1738.6$-$5333: $V_{\rm mag}$ from ESO image 
	
$^s$ WGAJ1808.2$-$5011: $V_{\rm mag}$ from ESO image 
	
$^t$ WGAJ1826.1$-$3650: $R_{\rm mag}$ from spectrum 
	
$^u$ WGAJ1911.8$-$2102: USNO magnitudes 
	
$^v$ WGAJ2151.3$-$4233: $\alpha_{\rm r}$ from PMN-Parkes data
	
$^w$ WGAJ2151.3$-$4233: magnitude from the APM Bright Galaxy Catalogue (Loveday 1996) 
	
$^x$ WGAJ2159.3$-$1500: $\alpha_{\rm ATCA} = 1.46$
	
$^y$ WGAJ2329.0$+$0834: magnitude from DSS 2 

\end{minipage}
\end{table*}

Of the 102 newly identified sources 86 are quasars (FSRQ and SSRQ) and
10 are BL Lacs. Adding the identifications published in Paper I, our
technique is $\approx 90\%$ efficient at selecting radio-loud quasars
and BL Lacs.

Nine objects in this publication are classified as radio
galaxies. Note that these are sources with a relatively flat radio
spectral index and therefore their jets are expected to be oriented
closer to the line of sight than the ones with steeper $\alpha_{\rm
r}$ ($> 0.7$), typical of lobe dominated radio galaxies.

Stars have not been detected in the DXRBS. This is due to their weak
radio emission (Helfand et al. 1999) with average fluxes below the
limits of the radio catalogues employed in our survey.

Classifications, redshifts and their uncertainties, along with other
observational properties for the objects presented here are given in
Table 5. Column (1) gives the WGA name of the source whereas columns
(2) and (3) give the {\it ROSAT} count rate and hardness ratio,
respectively.  The latter is the ratio between the count rates in the
hard ($0.87-2.0$ keV) and medium ($0.4-0.86$ keV) bands (see Padovani
\& Giommi 1996 for details).  Columns (4) and (5) give the \mbox{0.1 -
2.0 keV} X-ray flux (not corrected for galactic extinction) and the
unabsorbed X-ray flux at 1 keV respectively. Both X-ray fluxes have
been derived from the {\it ROSAT} count rates using the observed hardness
ratio and assuming galactic $N_{\rm H}$. For objects observed more
than once by {\it ROSAT}, we give the count rates, hardness ratios, and
X-ray fluxes found for each observation. Column (6) gives the ratio
between the $0.3 - 3.5$ keV X-ray flux and the radio flux at 6 cm,
columns (7) and (8) give the radio flux at 6 cm and radio spectral
index, the latter derived as described in \S~2.2.  Columns (9), (10)
and (11) give the magnitudes in the B$_j$ (from COSMOS), and O and E
(from APM) spectral bands. For most of the faint objects the magnitude
comes from other sources (see footnotes to table). The classification
and redshift of the object can be found in the last two columns, (12)
and (13). The redshift was computed, whenever possible, by taking the
mean of the values derived from the narrow lines. Where only a single
(broad) emission line was observed, it was assumed to be Mg II
$\lambda 2798$ \AA. When only weak features were present the redshift
is marked as questionable in the table and the object is discussed
individually. The redshift uncertainty represents the sample standard
deviation. No error on the redshift is listed if only one emission
line is present in the optical spectrum.

For objects which we classify as either radio galaxies or BL Lacs, we
give the Ca break strength and the rest frame equivalent width of the
strongest emission line in Table 6. Note that objects without a
determined redshift have been excluded from this table, as well as
WGAJ1120.4$+$5855, WGAJ1204.2$-$0710, and WGAJ1320.4$+$0140, for which the obtained
spectrum does not cover the region of the Ca break. The error
computation for the Ca break was based on the SNR blueward and redward
of this feature.

\vspace{0.2cm}

\begin{table}
{\bf Table 6.} Spectral Characteristics of BL Lacs and Radio\\
 Galaxies
\begin{center}
\begin{tabular}{lcrl} 
\hline 
Name & Ca break & EW [\AA] & Line \\
\hline
WGAJ0305.3$-$2420 & ? & 25$\pm$@3& H$\alpha$\\
WGAJ0421.5$+$1433 & $0.51\pm0.14$ & $\le$@6& \\
WGAJ0528.5$-$5820 & $0.38\pm0.18$ & 26$\pm$@4& [O II]\\
WGAJ0847.2$+$1133 & $\le0.05$ & $\le$@3&\\
WGAJ1056.9$-$7649 & $0.43\pm0.35$ & 32$\pm$@3& H$\alpha$\\
WGAJ1057.6$-$7724 & $0.48\pm0.18$ & $\le$@8 & \\
WGAJ1229.5$+$2711 & ? & 73$\pm$29 & [O II]\\
WGAJ1311.3$-$0521 & $0.33\pm0.12$ & $\le$@2 & \\
WGAJ1457.9$-$2124 & $0.45\pm0.20$ & 14$\pm$@2 & H$\alpha$\\
WGAJ1840.9$+$5452 & $0.06\pm0.12$ & 8$\pm$@1 & [O III]\\
WGAJ1936.8$-$4719 & $\le0.02$ & $\le$@3 & \\
WGAJ2151.3$-$4233 & $0.49\pm0.22$ & 6$\pm$@1 & H$\alpha$\\
WGAJ2330.6$-$3724 & $0.10\pm0.04$ & @2$\pm$@1& [O III]\\
\hline
\end{tabular}
\end{center}
\end{table}

\subsection{Notes on Individual Objects}

{\bf WGAJ0305.3$-$2420} The Ca break for this object is located in
noise and its measurement therefore impossible. Based on the relatively large
equivalent width of its strongest line, we classify it as a
radio galaxy, but note that a classification as a BL Lac cannot be
excluded.

{\bf WGAJ0421.5$+$1433} Based on a newly acquired spectrum with a
higher SNR than the one published in Paper I we determine the redshift
for this object ($z = 0.059\pm0.001)$ and reclassify it as a galaxy
based on a Ca break value of \mbox{0.51$\pm$0.14}. However, taking the
error of the Ca break into account, a classification of this object as
a BL Lac cannot be excluded.

{\bf WGAJ0513.8$+$0156} This is a problematic source. The (uncertain)
BL Lac classification of this object given in Paper I was based on
earlier information from the NVSS, which at that time listed only one
radio source at the position given in Paper I.  The error circle of
the GB6 now includes three NVSS sources. The object published in Paper
I is next to the brightest one at $z=0.084$. Further observations
yielded two other galaxies at redshifts $z = 0.087$ and $z = 0.092$,
the former close to the faintest NVSS source, the latter in between
the two brightest radio sources. The similar redshifts of the observed
galaxies suggest a small group. With the information available, it is
currently impossible to say which object is the radio counterpart of
the WGA source.

{\bf WGAJ0528.5$-$5820} We refine the redshift of this source to $z =
0.254\pm0.001$; however, the classification as a BL Lac is
unaffected. The spectrum of a star close to the DXRBS candidate has
been erroneously identified as the optical counterpart in Paper I.

{\bf WGAJ0847.2$+$1133} This object was identified independently by
Cao et al. (1999). We confirm its identification and redshift.

{\bf WGAJ0937.1$+$5008} This object was identified independently by
Henstock et al. (1997). We confirm its identification and redshift.
Our spectrum, however, is significantly bluer and reflects a higher
continuum level. As a result, the emission lines in our spectrum are
correspondingly weaker.

{\bf WGAJ1056.9$-$7649} We classify this object as a galaxy, but
given the large error on the Ca break we cannot exclude a
classification as a BL Lac.

{\bf WGAJ1057.6$-$7724} This object was classified in Paper I as a BL
Lac with an uncertain redshift of $z = 0.541$. A new spectrum obtained
in March 1999 at ESO 3.6 m yields a reclassification of this source as
a radio galaxy with a redshift of $z = 0.181\pm0.001$ and a Ca break
value of \mbox{0.48$\pm$0.18}. Note that the error in the contrast
puts this object close to the dividing value of 0.4 adopted by
March\~a et al. (1996) to distinguish radio galaxies from BL Lac
objects.

{\bf WGAJ1101.8$+$6241} This object was identified independently by
Henstock et al. (1997). We confirm its identification and redshift.

{\bf WGAJ1116.1$+$0828} The redshift for this object given in Table 5
refers to the emission line.  The absorption system detected is at
$z=0.163\pm0.001$.

{\bf WGAJ1120.4$+$5855} The classification of this object as a NLRG is
based on the rest-frame equivalent width of 40 \AA~for the strongest
emission line [O III] $\lambda 5007$ and is ambiguous without an
information on the Ca break strength.

{\bf WGAJ1204.2$-$0710} The claimed redshift for this object is tentative
due to the weakness of the identified absorption features. 

{\bf WGAJ1222.6$+$2934} A newly acquired spectrum of this source
allows a redetermination of the redshift to $z = 0.787\pm0.001$;
however, the classification as an FSRQ is unaffected.

{\bf WGAJ1320.4$+$0140} The redshift given in Drinkwater et
al. (1997) for this object coincides within the errors with the
redshift determined from our spectrum. The strongest emission line C
III] has a rest-frame equivalent width of 9 \AA, therefore we classify
this object as a BL Lac.

{\bf WGAJ1323.8$-$3653} We classify this faint object as a BL Lac,
based on the absence of strong emission and absorption
features. However, we cannot claim this object to be completely
featureless, in particular above $\lambda \sim 7,000$ \AA, because 
in that region the lower SNR renders the sky subtraction problematic.   

{\bf WGAJ1329.0$+$5009} This object was identified independently by
Falco, Kochanek, \& Mu\~noz (1998). We confirm its identification and
redshift.

{\bf WGAJ1420.6$+$0650} This source seems to be partially obscured.
We detect a broad H$\alpha$ emission line, but no H$\beta$ emission is
visible.

{\bf WGAJ1539.1$-$0658} We classify this object as a BL Lac, based on
the absence of strong emission and absorption features. However, we
cannot claim this object to be completely featureless, in particular
above $\lambda \sim 7,000$ \AA, because in that region the lower SNR 
renders the sky subtraction problematic. 

{\bf WGAJ1656.6$+$5321} This object was identified independently by
Falco, Kochanek, \& Mu\~noz (1998). We confirm its identification and
redshift.

\section{Sample Properties}

As of July 2000\footnote{We include here the spectra of two BL Lacs we
reobserved in August 2000, as those spectra are of better quality than 
those we previously had.}, 189 objects have been spectroscopically
identified. Of these, 151 (80.0\%) are quasars, 24 (12.6\%) are BL
Lacs, and 14 (7.4\%) are NLRG. These numbers include 10 sources which
have been observed as part of other surveys and whose classification
was made available to us before publication. Another 109 previously
known objects are also in the sample, of which 83 are quasars, 12 are
BL Lacs, and 14 are NLRG. These numbers are larger than those reported
in Paper I because of our expanded criteria, discussed in \S~2. Our
sample now contains 234 quasars (181 FSRQ and 53 SSRQ), 36 BL Lacs,
and 28 NLRG, for a total of 298 identified sources. As a result of the
observations presented here we have increased the number of newly
identified objects in our sample by more than 50\%. Table 7 summarizes
the status of the identifications.

\begin{table*}
{\bf Table 7.} Sample Identification Status\\
\begin{center}
\begin{tabular}{lccccccc} 
\hline 
Class & \multicolumn{2}{c}{Newly} & Previously & \multicolumn{2}{c}{Total} & Uncertain & No z \\
& \multicolumn{2}{c}{identified} & known & & & Class & \\
\hline
Radio-Loud Quasars  & 151 & 139 & 83 & 234 & 222 & 0 & @0 \\
BL Lacs        &  @24 & @19 & 12 & @36 & @31 & 3 & 11 \\ 
Radio Galaxies &  @14 & @@8& 14 & @28 & @22 & 4 & @0 \\
\hline
Total          & 189 & 166 & 109 & 298 & 275 & 7 & 11 \\ 
\hline
\end{tabular}
\end{center}
The numbers to the right in columns (2) and (4) refer to sources which fulfill all DXRBS 
selection criteria.
\end{table*}

The subsample of sources which fulfill all our
selection criteria [\S~2 and Paper I] and which will be used for the
detailed analysis is now 85\% identified. It includes 275 objects,
of which 222 are quasars (175 FSRQ and 47 SSRQ), 31 are BL Lacs, and
22 are NLRG. An ESO Very Large Telescope (VLT) program to identify 15
very faint ($R \sim 23$) DXRBS candidates has been partially completed 
(data reduction is in progress); another such program is currently 
under way. Approximately 50 sources remain to be
identified (about half of them have been observed during our ESO run in
August 2000; data reduction is in progress). These sources are about 
one magnitude fainter than the
rest of the DXRBS sample ($\langle V \rangle \sim 19.2$ as compared to
18.3). Based on our results so far, we expect them to be mostly
quasars at relatively high ($z \ga 1.5$) redshifts. With this in mind,
we now review the properties of the sample.

\subsection{Redshift Distributions}

Figure 2 displays the fractional redshift distribution of the DXRBS
quasars compared to the quasars found in the S4 ($f_{\rm 5GHz} > 0.5$
Jy) and 1 Jy ($f_{\rm 5GHz} > 1$ Jy) surveys (Stickel \& K\"uhr 1994;
Stickel, Meisenheimer \& K\"uhr 1994). Note that DXRBS quasars have
been selected to have $\alpha_r\le0.7$. Therefore the S4 and 1Jy
distributions shown here comprise the quasars from these samples
defined on the same basis. 

While for the S4 and 1 Jy surveys the area of the sky covered is the
same at all fluxes (5,624 and 32,208 deg$^2$ respectively), this is
not the case for DXRBS. Due to the serendipitous nature of the survey
and the variable sensitivity of the {\it ROSAT} PSPC
detector across the field of view, the area in which faint X-ray
sources could be detected is smaller than that for brighter X-ray
sources. The DXRBS redshift distribution has therefore been de-convolved
with the appropriate sky coverage and this is what is shown in
Fig. 2. Namely, each bin represents $\sum 1/{\rm Area}(f_{\rm x})$ for
all the sources in that bin, where Area$(f_{\rm x})$ is the area accessible
at its X-ray flux, divided by the total surface density of sources.    
Error bars represent the $1\sigma$ range based on Poisson statistics. 
We note that the sky coverage is difficult to determine in the
parts of the PSPC field of view affected by the rib structure
($13^{\prime} < {\rm offset} < 24^{\prime}$). That area, and the
sources within, have therefore been excluded from the analysis
presented in this subsection. Moreover, only sources with $f_{\rm
5GHz} > 51$ mJy have been included since we still not have computed
the sky coverage of the PMN survey below this flux.

As the radio flux limit of the quasar samples drops from 1 Jy to $\sim
0.05$ Jy, the approximate limit of the DXRBS, a progression towards
higher redshifts is clearly seen, as expected. The mean redshift, in
fact, moves from $\langle z \rangle = 1.18\pm0.05$ for the 1 Jy, to
$\langle z \rangle = 1.30\pm0.07$ for the S4, to $\langle z \rangle =
1.56\pm0.06$ for the DXRBS sample. A Kolmogorov-Smirnov test shows
that the DXRBS redshift distribution is significantly different ($P >
99\%$) from the 1 Jy and S4 distributions. The mean values are also
different at the same significance level. Considering that the DXRBS
quasar sample is $\sim 85\%$ complete and that the missing sources are
likely to be at relatively high ($z \ga 1.5$) redshift, we expect the
final mean redshift to be even higher.   
Compared to the S4 and 1 Jy samples, a larger fraction of
DXRBS quasars are at relatively high redshift.  In fact, DXRBS finds
$\sim 30\%$ of its quasars at $z > 2$ (once the effect of the WGACAT
sky coverage is taken into account), whereas only $\sim 15\%$ of the
S4 and 1 Jy quasars lie above this redshift.

\setcounter{figure}{1}
\begin{figure}
\centerline{\psfig{figure=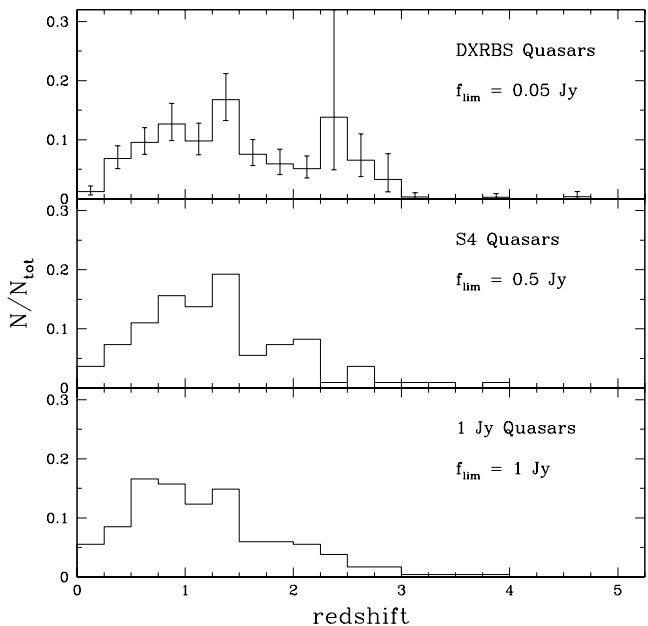,width=9cm}}
\caption{Fractional redshift distribution for the 148 DXRBS, 109 S4, and 
235 1 Jy quasars
with $\alpha_{\rm r} \le 0.7$. The DXRBS distribution 
has been de-convolved with the appropriate sky coverage. Error bars represent 
the $1\sigma$ range based on Poisson statistics.}
\end{figure}

\setcounter{figure}{2}
\begin{figure}
\centerline{\psfig{figure=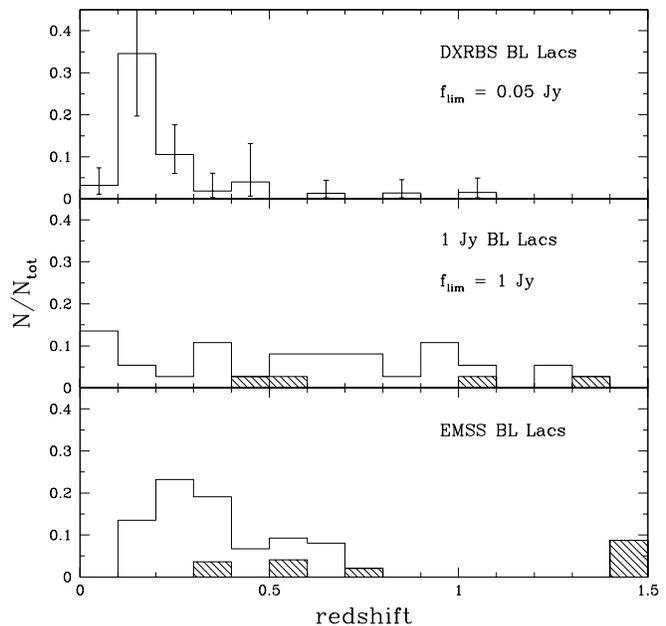,width=9cm}}
\caption{Fractional redshift distribution for the 17 DXRBS, 32 1 Jy, and 38 
EMSS BL Lacs. The DXRBS and EMSS distributions have been de-convolved with 
the appropriate sky coverages. The hatched areas represent lower limits (1 Jy) 
and uncertain values (EMSS). Error bars represent the $1\sigma$ range based 
on Poisson statistics. See text for details.} 
\end{figure}

Figure 3 displays the redshift distribution of the DXRBS BL Lacs,
compared to that of the 1 Jy (Stickel et al. 1991; Stickel \& K\"uhr
1994; Stocke \& Rector 1997) and EMSS (Rector et al. 2000)
samples. The DXRBS and EMSS redshift distributions have been
de-convolved with the appropriate sky coverage (the sky coverage for
the EMSS was taken from Gioia et al. 1990 and Morris, private
communication). We have not included BL Lacs from the {\it Einstein}
Slew Survey (Perlman et al. 1996a) in this plot because the sky
coverage of that survey is not fully known yet. Five EMSS redshifts
are uncertain, while four 1 Jy redshifts are lower limits (hatched
areas in Fig. 3; five more 1 Jy sources have a lower limit on their
redshift of 0.2 based on the non-detection of their host galaxies on
the optical image). Note that the fraction of BL Lacs with redshift
information ranges from $93\%$ and $86\%$ for the EMSS and 1 Jy
samples respectively to $63\%$ for DXRBS. Half of the DXRBS BL Lacs
without a determined redshift are previously known objects, and
therefore have not been spectroscopically observed by us.

The mean redshift for the three BL Lac samples is 0.24 for DXRBS, 0.46
for the EMSS, and 0.63 (including lower limits) for the 1 Jy. The
DXRBS and EMSS samples are peaked at $z=0.2$ and $z=0.3-0.4$
respectively, and neither sample has significant numbers of $z>0.8$
objects ($\sim 9\%$ in the EMSS and $\sim 12\%$ in the DXRBS). By
comparison, the 1 Jy BL Lacs have an essentially flat redshift
distribution out to nearly $z=1.5$, with 10/32 1 Jy BL Lacs at $z >
0.8$ and 5 at $z>1$.  In contrast to the quasars, no progression to
higher redshifts is seen for the BL Lacs as the radio flux limit drops
from 1 Jy, for the 1 Jy sample, to 0.05 Jy for the DXRBS. Similarly,
no progression of this kind is seen as the X-ray flux limit drops from
a few $\times 10^{-13}$ erg cm$^{-2}$ s$^{-1}$ for the EMSS to a few
$\times 10^{-14}$ erg cm$^{-2}$ s$^{-1}$ for the DXRBS. In this paper,
however, we will not attempt a detailed comparison of the three BL Lac
redshift distributions. At this stage, this would not be meaningful,
first because, as mentioned above, about one third of the DXRBS BL
Lacs are not included in Figure 3 since they lack redshifts, and
second because the identification of DXRBS is not $100\%$ complete
yet. We would only like to note that the criteria employed to classify
objects as a BL Lac were somewhat different in each of these
surveys. As we previously pointed out in Paper I, while the
spectroscopic criteria employed for DXRBS and the EMSS were identical
(once the EMSS criteria were expanded to take into account the
March\~a et al. (1996) expansion of the original BL Lac region in the
$(C,W_\lambda)$ plot; see Rector et al. 1999 for discussion), the EMSS
did not require a flat radio spectrum to classify an X-ray source as a
candidate BL Lac. The 1 Jy survey used a different set of criteria to
classify objects as BL Lacs, requiring $\alpha_r\le0.5$ and
$W_\lambda<5(1+z)$ \AA~but with no restriction on the Ca break. In
addition, as was detailed in Paper I and Perlman et al. (1996b), the 1
Jy survey was often inconsistent in applying its selection criteria.
Similar points have also been made by other authors (March\~a \&
Browne 1995; Stocke \& Rector 1997; Rector, Stocke \& Perlman 1999).

We note the striking difference in redshift range between BL Lacs and
quasars, the latter reaching $z\sim 4.7$, the former only getting up
to $z\sim 1.5$. Note that, while $\sim 37\%$ of DXRBS BL Lacs lack a
redshift determination, that is not the case for the EMSS and 1 Jy
samples for which redshifts are available for $\sim 90\%$ of the
sources. Although one cannot exclude the possibility that the missing
redshifts are all relatively high, the point remains that the mean
redshift values for BL Lacs and quasars {\it belonging to the same
sample} are significantly different.

Finally, the redshift distribution for the radio galaxies in our
sample reaches only $z \sim 0.4$ and is skewed towards low redshifts,
with $\langle z \rangle = 0.13$.

\subsection{X-ray/Radio Luminosities}

Because it reaches much lower fluxes than any previously published
sample of FSRQ (by a factor $10-20$ in radio flux and by a factor 10
in X-ray flux), DXRBS has vastly expanded the coverage of parameter
space for FSRQ in complete samples. These expansions occur in two
regions of the $L_{\rm x},L_{\rm r}$ plane: (1) at low luminosities;
(2) at high $L_{\rm x}/L_{\rm r}$ values, with the discovery of a new
class of X-ray bright FSRQ.

At the low luminosity end, DXRBS includes at present $\sim 5 \times$
more FSRQ with $L_{\rm r} < 10^{26.5} {\rm ~W ~Hz^{-1}}$ than the S4
and 1 Jy samples combined. About 40 DXRBS FSRQ are in fact already
known at these low powers, and this will put strong constraints on the
FSRQ radio luminosity function (LF). For comparison, note that the
radio LF derived by Urry \& Padovani (1995) from the 2 Jy sample (the
only published high frequency sample with relatively complete redshift
information) included only one source at $L_{\rm r} < 10^{26.5} {\rm
~W ~Hz^{-1}}$, a power which coincides roughly with the predicted
flattening of the LF based on unified schemes. Five of these low-power 
objects are in a luminosity range never
before reached for FSRQ: $10^{24.5} < L_{\rm r} < 10^{25.5}$ W Hz$^{-1}$, 
the former
being the expected minimum luminosity of FSRQ assuming these objects
are beamed FR II radio galaxies.

Paper I discussed the finding of a large fraction ($\sim 25\%$) of
X-ray bright ($\log L_{\rm x}/L_{\rm r} > -6$ or $\alpha_{\rm rx} <
0.78$) FSRQ in the DXRBS, and termed them high-energy peaked FSRQ
(HFSRQ), stressing the similarity in their broadband spectral shapes
to the high-energy peaked BL Lacs. The fraction of HFSRQ in the DXRBS
is now $28\%$. We also find 14 high-energy peaked SSRQ.  By
comparison, the previously known complete samples included very few of
these objects ($3\%$ in total). As discussed in Paper I, the reason
for this discrepancy is because DXRBS is the first X-ray survey which
included the radio spectral slope in its identification process, thus
for the first time allowing the identification of a sample of X-ray
emitting FSRQ. Importantly, further research on other samples has
shown that the X-ray/radio luminosity distribution of DXRBS FSRQ is
not anomalous. A slightly larger percentage of these objects is found
among the FSRQ in the RGB survey (Laurent-Muehleisen et al. 1998,
Padovani et al., in preparation), which goes to somewhat larger values
of $L_{\rm x}/L_{\rm r}$ but at higher X-ray flux than
DXRBS. Moreover, subsequent archival research on the {\it Einstein}
EMSS and Slew samples (Perlman et al. 2000) has also shown similar
fractions of HFSRQ.

As regards BL Lacs, DXRBS will be providing the first complete
radio-selected sample down to $\sim 50$ mJy, an improvement of a
factor of 20 in flux over the 1 Jy sample, still the only sizeable BL
Lac radio-selected sample. The similar improvement in radio powers
will allow us to extend the radio LF of radio-selected BL Lacs down to
$L_{\rm r} \sim 10^{23-24} {\rm ~W ~Hz^{-1}}$, again getting close to
the expected minimum powers according to unified schemes (Urry \&
Padovani 1995). Moreover, due to their low X-ray fluxes, DXRBS BL Lacs
are also reaching more than an order of magnitude lower $L_{\rm x}$
than, for example, the EMSS or Slew samples.

\subsection{A Multiwavelength Picture of DXRBS}

A broader picture of the DXRBS sample is shown in Fig. 4, which plots
the $\alpha_{\rm ro}$, $\alpha_{\rm ox}$ plane for DXRBS
sources. These are the usual effective spectral indices defined
between the rest-frame frequencies of 5 GHz, 5,000 \AA, and 1
keV. X-ray and optical fluxes have been corrected for Galactic
absorption. The effective spectral indices have been k-corrected using
the appropriate radio and X-ray spectral indices, available for each
source. Optical fluxes at 5,000 \AA~have been derived and k-corrected
as described in Padovani et al. (in preparation). The dashed lines
represent, from top to bottom, the loci of $\alpha_{\rm rx} = 0.85$,
typical of 1 Jy quasars with $\alpha_{\rm r} \le 0.70$ and BL Lacs,
$\alpha_{\rm rx} = 0.78$, the dividing line between HBL and LBL, and
$\alpha_{\rm rx} = 0.70$, the upper boundary for EMSS BL Lacs.

A few points are worth noticing in the figure: 
\begin{enumerate}

\item the majority of DXRBS BL Lacs occupy a somewhat intermediate
area in $\alpha_{\rm rx}$ space, in between those occupied by the 1 Jy
and EMSS BL Lac samples. Such objects are also being found in other
{\it ROSAT}-based surveys, for example RGB (Laurent-Muehleisen et al. 1998,
1999), RC (Kock et al. 1996) and HQS/RASS (Nass et al. 1996). It is
now clear that the bimodal distribution observed for the 1 Jy and
EMSS samples (see, e.g., Fig. 12 of Padovani \& Giommi 1995) was due
to selection effects and that deeper surveys are now filling the gap
in parameter space. Note that due to our X-ray and radio flux limits
only a handful of DXRBS BL Lacs have effective spectral indices
typical of classical HBL (e.g., the EMSS sample).

\item DXRBS quasars have significantly lower ($P>99.99\%$)
$\alpha_{\rm rx}$ than 1 Jy quasars: $\alpha_{\rm rx} = 0.82\pm0.01$
versus $0.86\pm0.01$. This corresponds to an X-ray-to-radio luminosity
ratio on average a factor of 2 larger for DXRBS. As discussed above,
DXRBS, by reaching relatively deep radio fluxes, is sampling the low
$\alpha_{\rm rx}$ (high $L_{\rm x}/L_{\rm r}$) end of the quasar
distribution. In particular, $28\%$ of DXRBS FSRQ have $\alpha_{\rm
rx}$ values typical of HBL, as compared to only $3\%$ for the 1 Jy
sample.

\item NLRG are concentrated towards the bottom right-hand corner of
the diagram and are characterized by relatively low $\alpha_{\rm ro}$
values ($\sim 0.4$) and relatively high $\alpha_{\rm ox}$ values
($\sim 1.7$). This is due to the strong thermal (stellar) component
present in these sources. Note that although some BL Lacs are also
present in this region, the mean $\alpha_{\rm ox}$ values for the two
classes ($1.38\pm0.06$ and $1.69\pm0.09$ for BL Lacs and radio
galaxies respectively) are significantly different ($P = 99.6\%$).

\item still unclassified sources are, on average, fainter in all bands than
identified ones but mostly so in the X-rays. This explains the fact that they
occupy the high $\alpha_{\rm rx}$ part of the $\alpha_{\rm ro}$, $\alpha_{\rm
ox}$ plane, the region typical of extreme FSRQ.  

\end{enumerate}

\setcounter{figure}{3}
\begin{figure}
\centerline{\psfig{figure=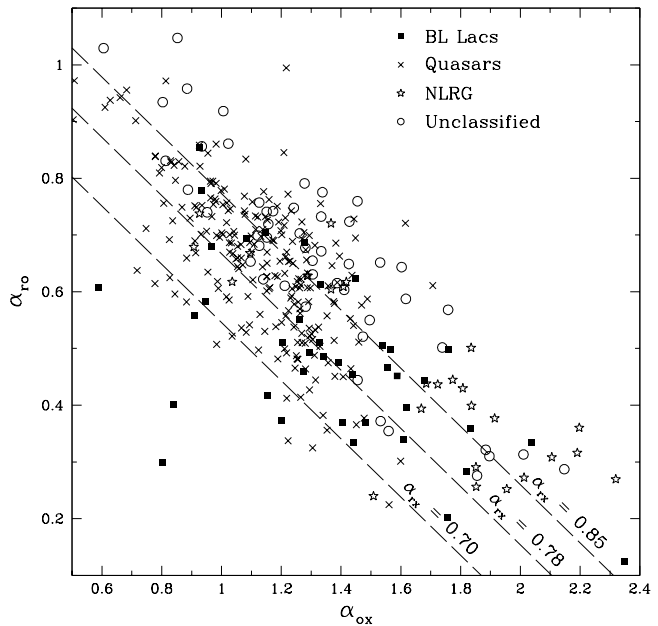,width=9cm}}
\caption{The $\alpha_{\rm ro}$, $\alpha_{\rm ox}$ plane for DXRBS sources.
BL Lac objects are shown as filled squares, quasars as crosses, NLRG as
stars, while still unidentified sources are shown as open circles. 
Effective spectral indices are defined the usual way and calculated between
the rest-frame frequencies of 5 GHz, 5000 \AA, and 1 keV. The dashed lines 
represent, from top to bottom, the loci of $\alpha_{\rm rx} = 0.85$, typical
of 1 Jy quasars with $\alpha_{\rm r} \le 0.70$ and BL Lacs, 
$\alpha_{\rm rx} = 0.78$, the dividing line
between HBL and LBL, and $\alpha_{\rm rx} = 0.70$, the upper boundary
for EMSS BL Lacs.}
\end{figure}

The impact of these findings, both for BL Lac and FSRQ, on our
multiwavelength picture of the blazar class will be discussed in a future
paper.

\section{Conclusions} 

DXRBS takes advantage of three of the most dominant blazar properties,
namely relatively strong X-ray and radio emission, and a flat radio
spectrum, to select a large sample of flat-spectrum radio quasars and
BL Lacs. Reaching 5 GHz radio fluxes $\sim 50$ mJy and $0.1-2.0$ keV
X-ray fluxes a few $\times 10^{-14}$ erg cm$^{-2}$ s$^{-1}$, DXRBS is
the faintest and largest flat-spectrum radio sample available to date
mapping previously unexplored regions of parameter space.

As of July 2000, our sample is $\sim 85\%$ identified and contains 298
sources, of which 234 (79\%) are quasars, 36 (12\%) are BL Lacs and 28
(9\%) are radio galaxies. Thus our technique has an efficiency of
$\sim 90\%$ at finding radio-loud quasars and BL Lacs. We have a
measured redshift for 96\% of our identified objects.

Our main results are the following: 

\begin{enumerate}
\item The DXRBS finds a large fraction of quasars at high redshift:
30\% of DXRBS quasars have $z > 2$ (once the effect of the WGACAT sky
coverage is taken into account), as compared to 15\% of the quasars
with $\alpha_{\rm r} \leq 0.70$ in the 1 Jy and S4 surveys combined.

\item The DXRBS has vastly expanded coverage of the faint end of the
radio luminosity function both for BL Lacs and FSRQ. Forty of our 181
DXRBS FSRQ are at $L_{\rm r} < 10^{25.5}$ W Hz$^{-1}$, compared to
only 10 out of 300 in the 1 Jy and S4 samples combined. Five of these
are in the luminosity range $10^{24.5} < L_{\rm r} < 10^{26.5}$ W
Hz$^{-1}$, the former being the expected minimum luminosity of FSRQ
assuming these objects are beamed FR II radio galaxies.

As regards BL Lacs, DXRBS will be providing the first complete
radio-selected sample down to $\sim 50$ mJy, an improvement of a
factor of 20 in flux over the 1 Jy sample, still the only sizeable BL
Lac radio-selected sample. The similar improvement in radio powers
will allow us to extend the radio LF of radio-selected BL Lacs down to
$L_{\rm r} \sim 10^{23-24} {\rm ~W ~Hz^{-1}}$, again getting close to
the expected minimum powers according to unified schemes.

\item DXRBS has filled large holes in the coverage of the $(L_{\rm x},
L_{\rm r})$ parameter space, both for quasars and BL Lacs. Prior to
DXRBS, basically no FSRQ within complete samples were known at values
of $L_{\rm x}/L_{\rm r} > 10^{-6}$, while 28\% of FSRQ in our sample
fall in this category. The discovery of such objects raises new
challenges for our understanding of blazar physics.

DXRBS contains a large number of BL Lacs with $10^{-6.5}<L_{\rm
x}/L_{\rm r}<10^{-5.5}$, that is with spectral shapes ``intermediate''
between the ``classical'' radio- and X-ray selected sources. Such
objects are also being found in other {\it ROSAT}-based surveys, which shows
that the bimodal distribution observed for the 1 Jy and EMSS samples
was simply due to selection effects.
\end{enumerate}

The fraction of identified sources in the DXRBS sample is now large
enough to allow us to study its evolutionary properties, luminosity
function, and implications for unified schemes at low powers. We will
address these issues in future publications.

\section*{Acknowledgements}

This paper is based on observations collected at the European Southern
Observatory, La Silla, Chile (proposals ESO N. 60.B-0313, 61.B-0288,
and 62.P-0257), Kitt Peak National Observatory, and the Australia
Telescope Compact Array, Narrabri, Australia. H.L. acknowledges
financial support from the Deutscher Akademischer Austauschdienst
(DAAD). P.P., E.P., and H.L. acknowledge ESO and KPNO personnel for
their assistance during the observing runs. E.P.  acknowledges ATNF
personnel, especially John Reynolds, Robin Wark and Henrietta May, for
their assistance before, during, and after ATCA observing runs. We wish
to thank Dave Jauncey and Sandra Savaglio for a careful reading of the
manuscript and Tommaso Treu for obtaining two of our spectra at the
ESO 3.6 m telescope in August 1999.  We acknowledge Isobel Hook, Peter
Shaver, Meg Urry, and John Stocke for useful discussions and Isobel
Hook and Anna Wolter for providing us with the identifications of some
sources prior to publication.  This research has made use of the
BROWSE program developed by the ESA {\it EXOSAT} observatory and by
NASA HEASARC, of the NASA/IPAC Extragalactic Database (NED), which is
operated by the Jet Propulsion Laboratory, California Institute of
Technology, under contract with the National Aeronautics and Space
Administration, and of the STScI Digitized Sky Survey, the ROE/NRL
COSMOS service, and the RGO/APM service. The Digitized Sky Surveys
were produced at the Space Telescope Science Institute under
U.S. Government grant NAG W-2166. The images of these surveys are
based on photographic data obtained using the Oschin Schmidt Telescope
on Palomar Mountain and the UK Schmidt Telescope. The plates were
processed into the present compressed digital form with the permission
of these institutions. The UK Schmidt Telescope was operated by the
Royal Observatory Edinburgh, with funding from the UK Science and
Engineering Research Council (later the UK Particle Physics and
Astronomy Research Council), until 1988 June, and thereafter by the
Anglo-Australian Observatory. The blue plates of the southern Sky
Atlas and its Equatorial Extension (together known as the SERC-J), as
well as the Equatorial Red (ER), and the Second Epoch [red] Survey
(SES) were all taken with the UK Schmidt.

\section*{Appendix}

We present here the optical spectra for the 106 DXRBS sources discussed
in this paper. The wavelength in \AA~is plotted on the x-axis while the 
y-axis gives the flux $f_\lambda$ in units of $10^{-17}$ erg 
cm$^{-2}$ s$^{-1}$ \AA$^{-1}$. 

\begin{figure*}
\psfig{figure=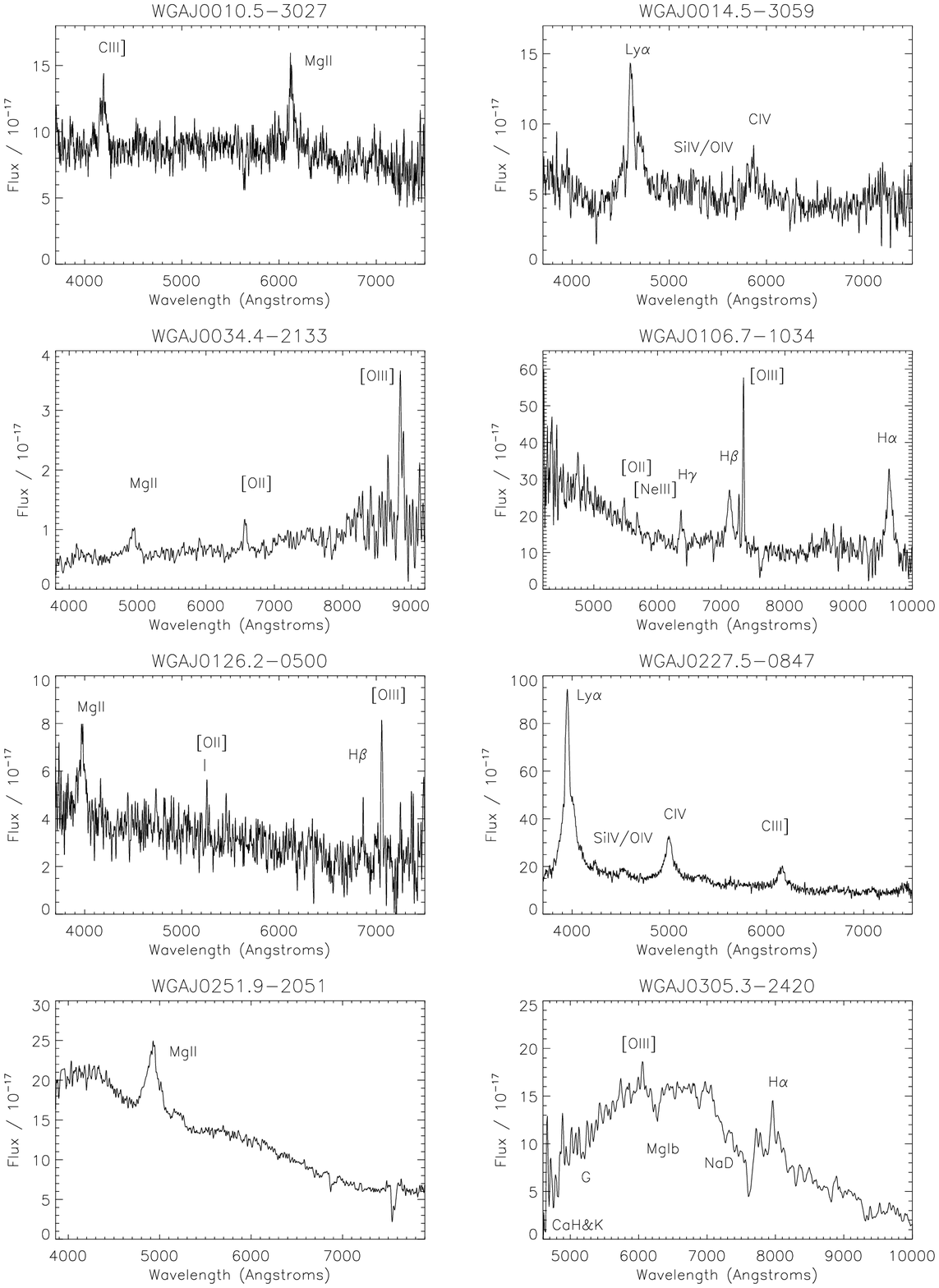,height=22.5cm} 
\end{figure*}

\begin{figure*}
\psfig{figure=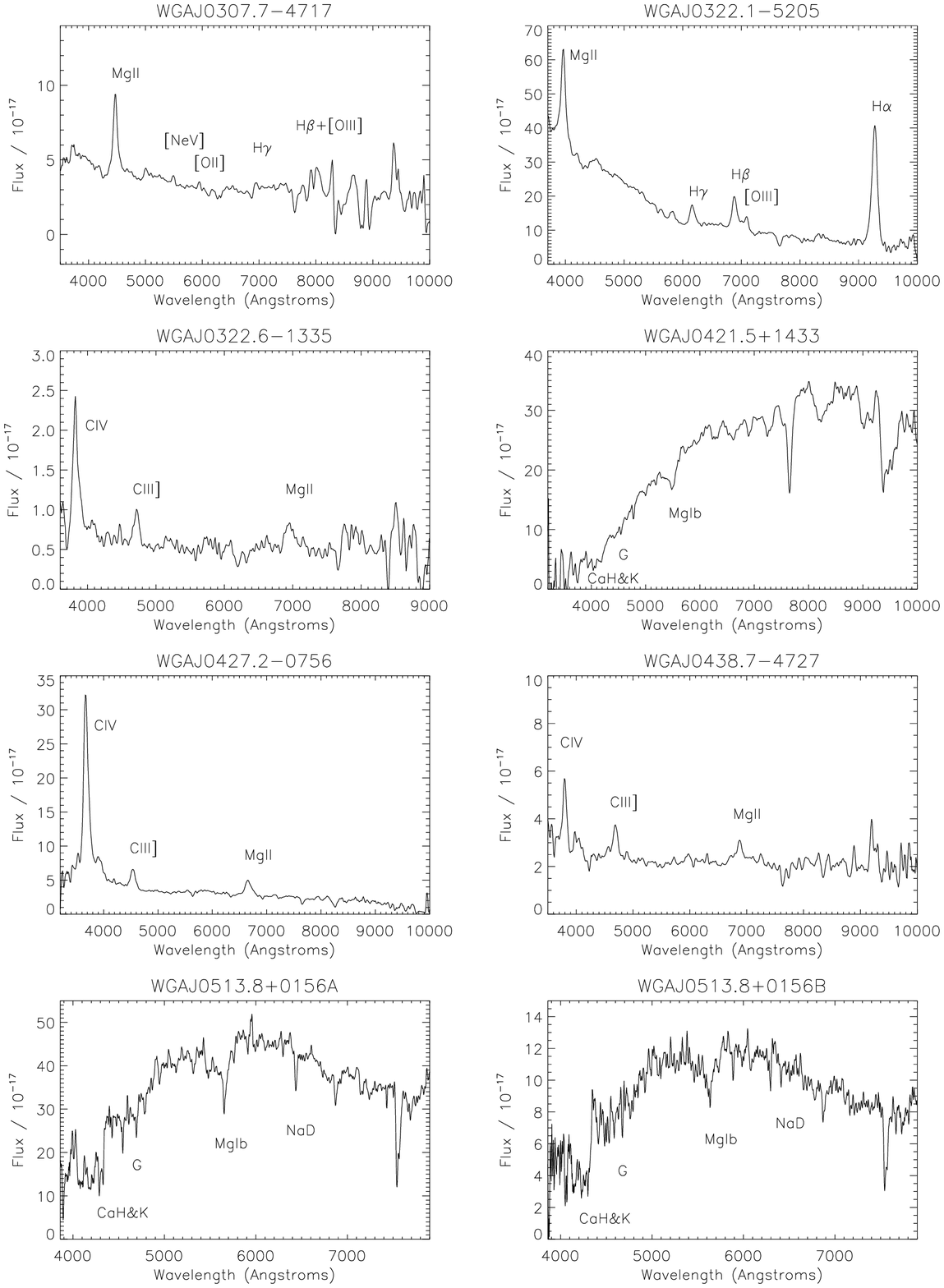,height=22.5cm}       
\end{figure*}

\begin{figure*}
\psfig{figure=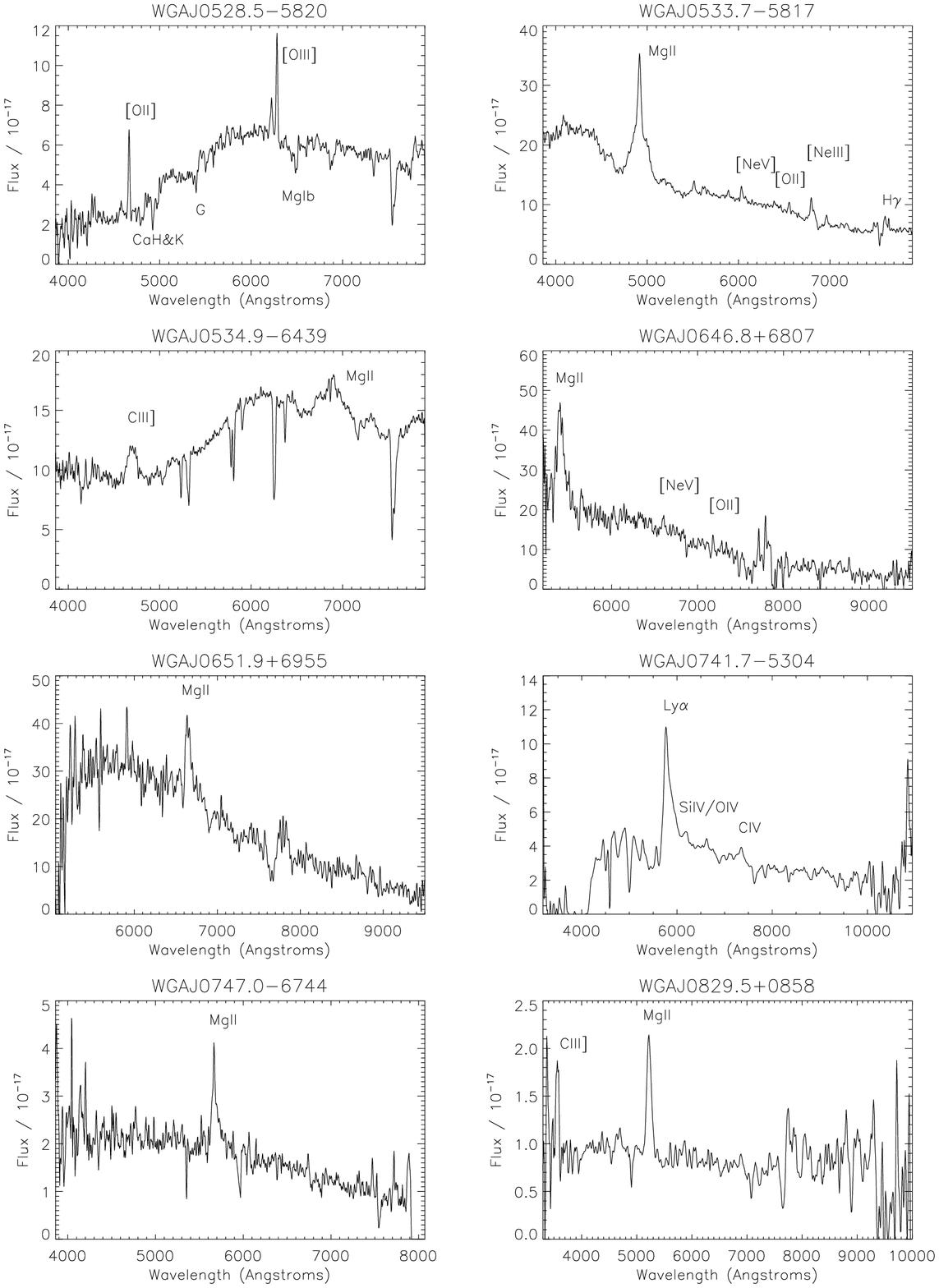,height=22.5cm}       
\end{figure*}

\begin{figure*}
\psfig{figure=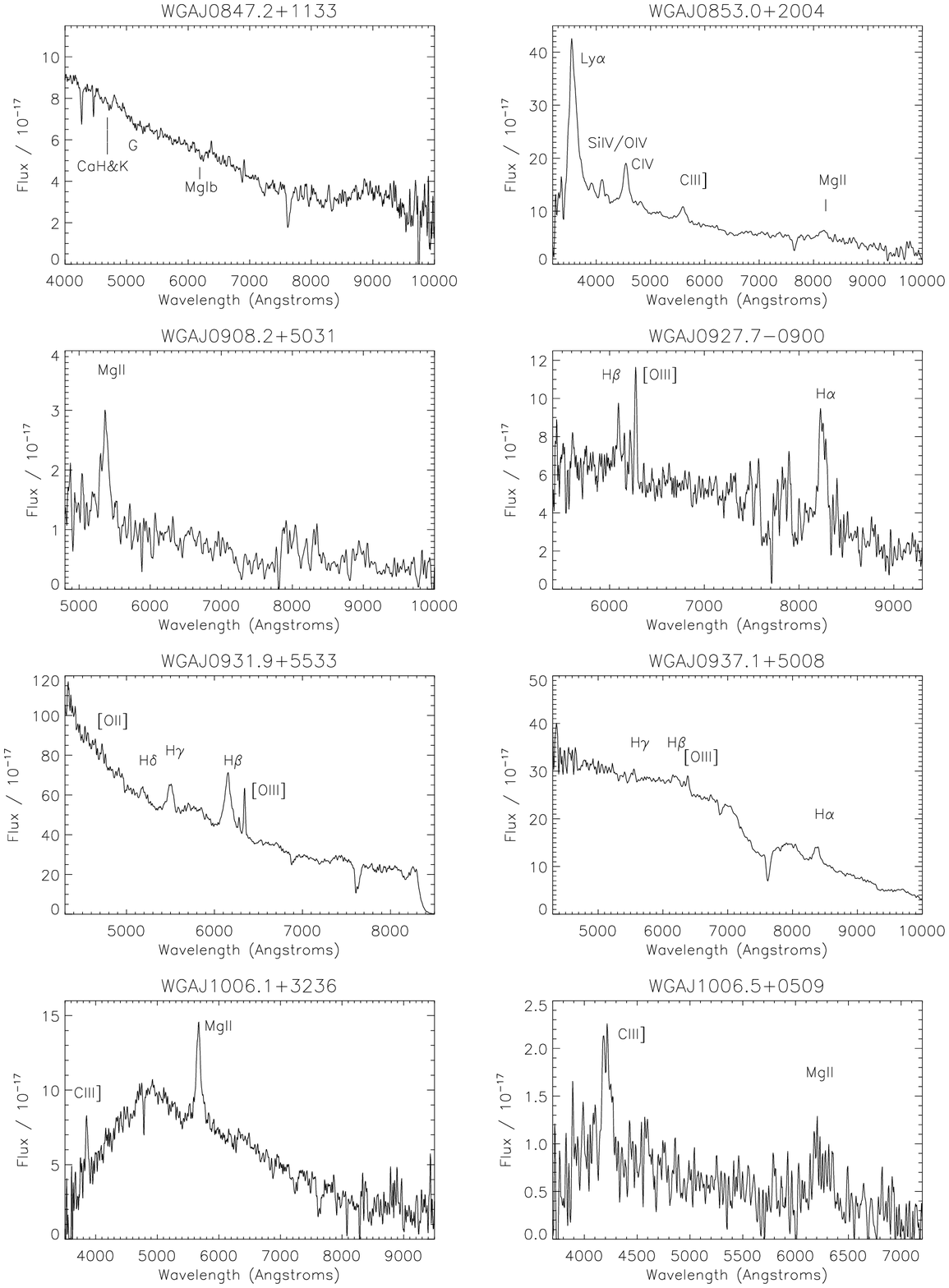,height=22.5cm}       
\end{figure*}

\begin{figure*}
\psfig{figure=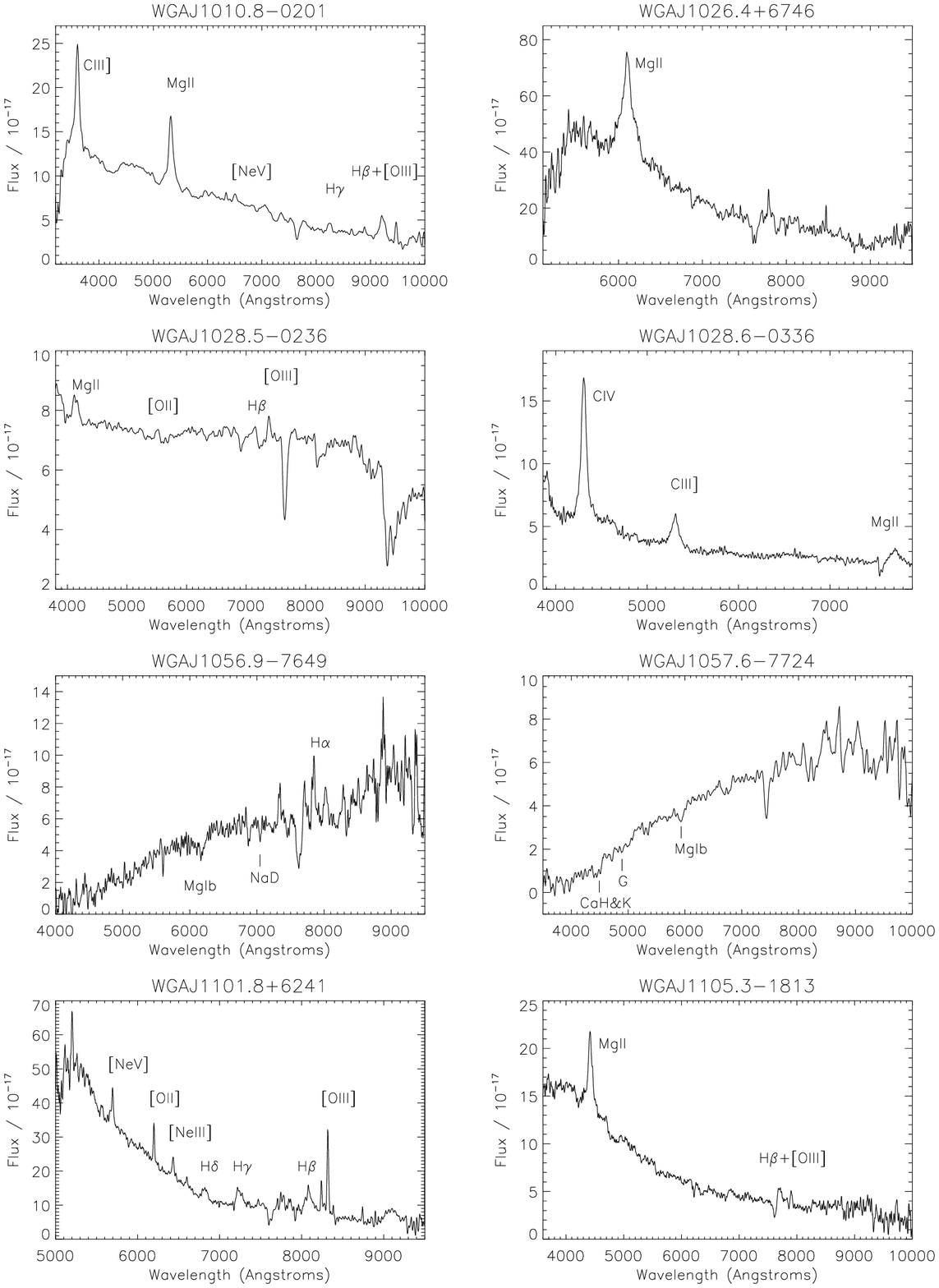,height=22.5cm}       
\end{figure*}

\begin{figure*}
\psfig{figure=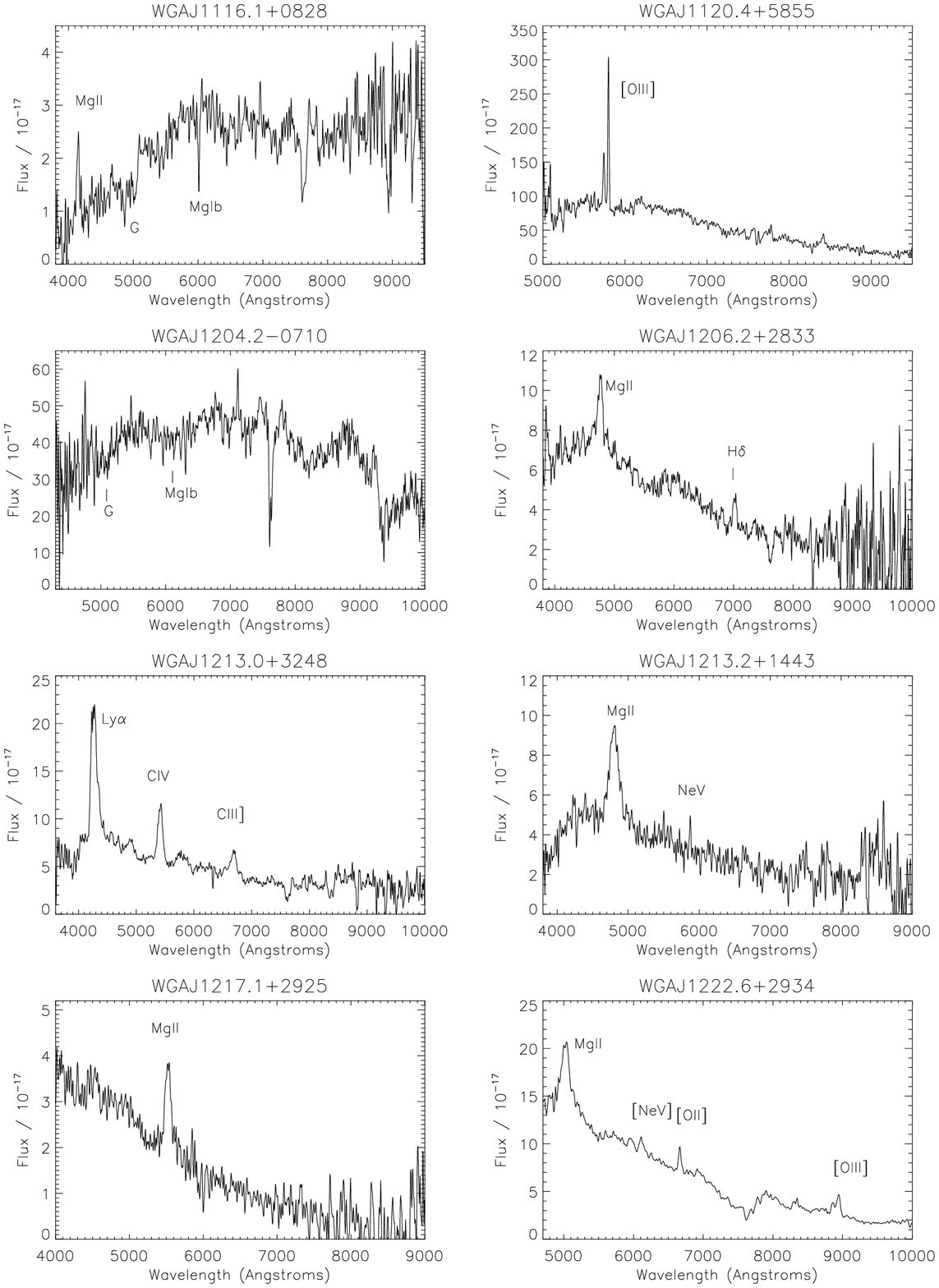,height=22.5cm}       
\end{figure*}

\begin{figure*}
\psfig{figure=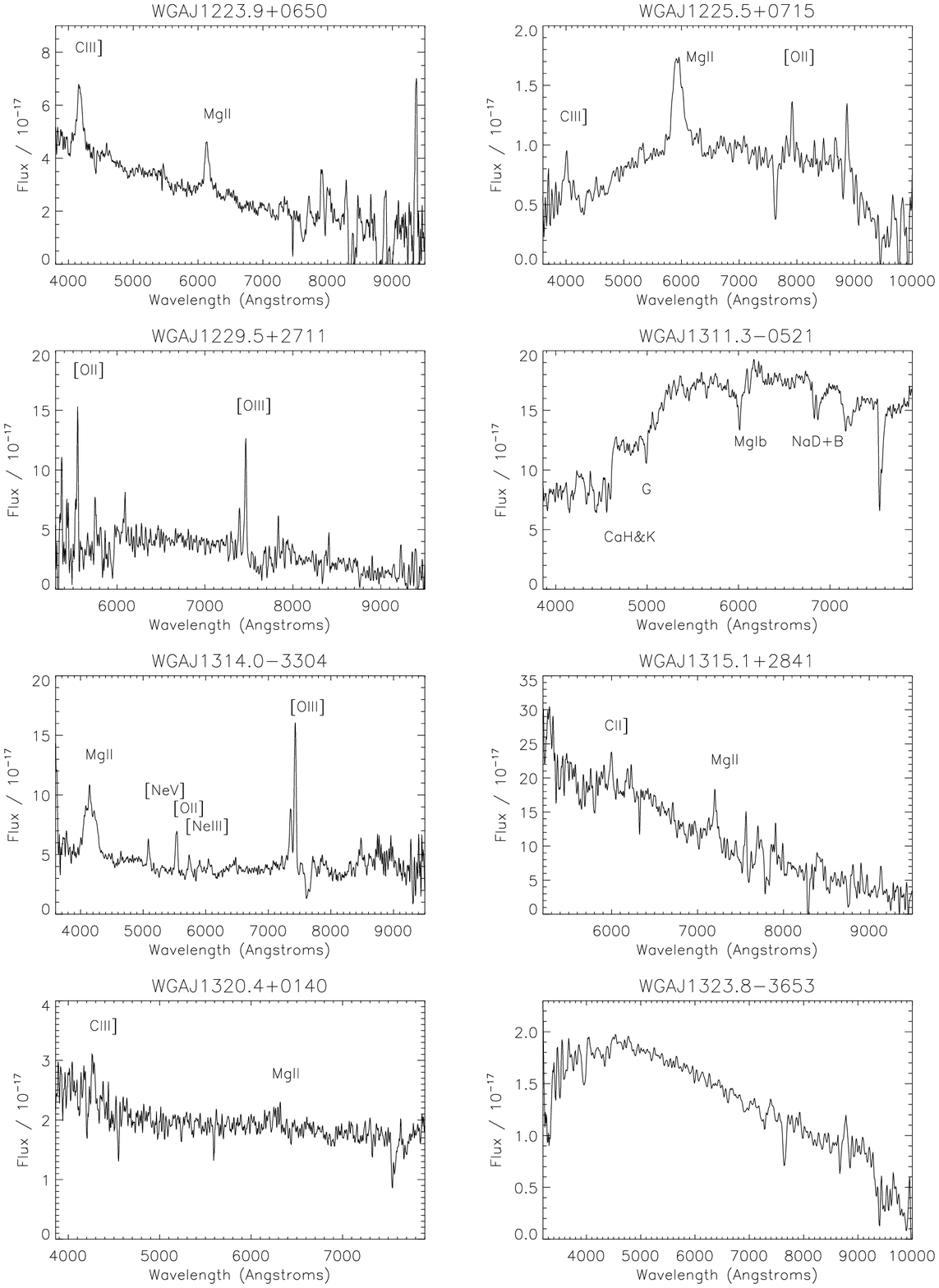,height=22.5cm}       
\end{figure*}

\begin{figure*}
\psfig{figure=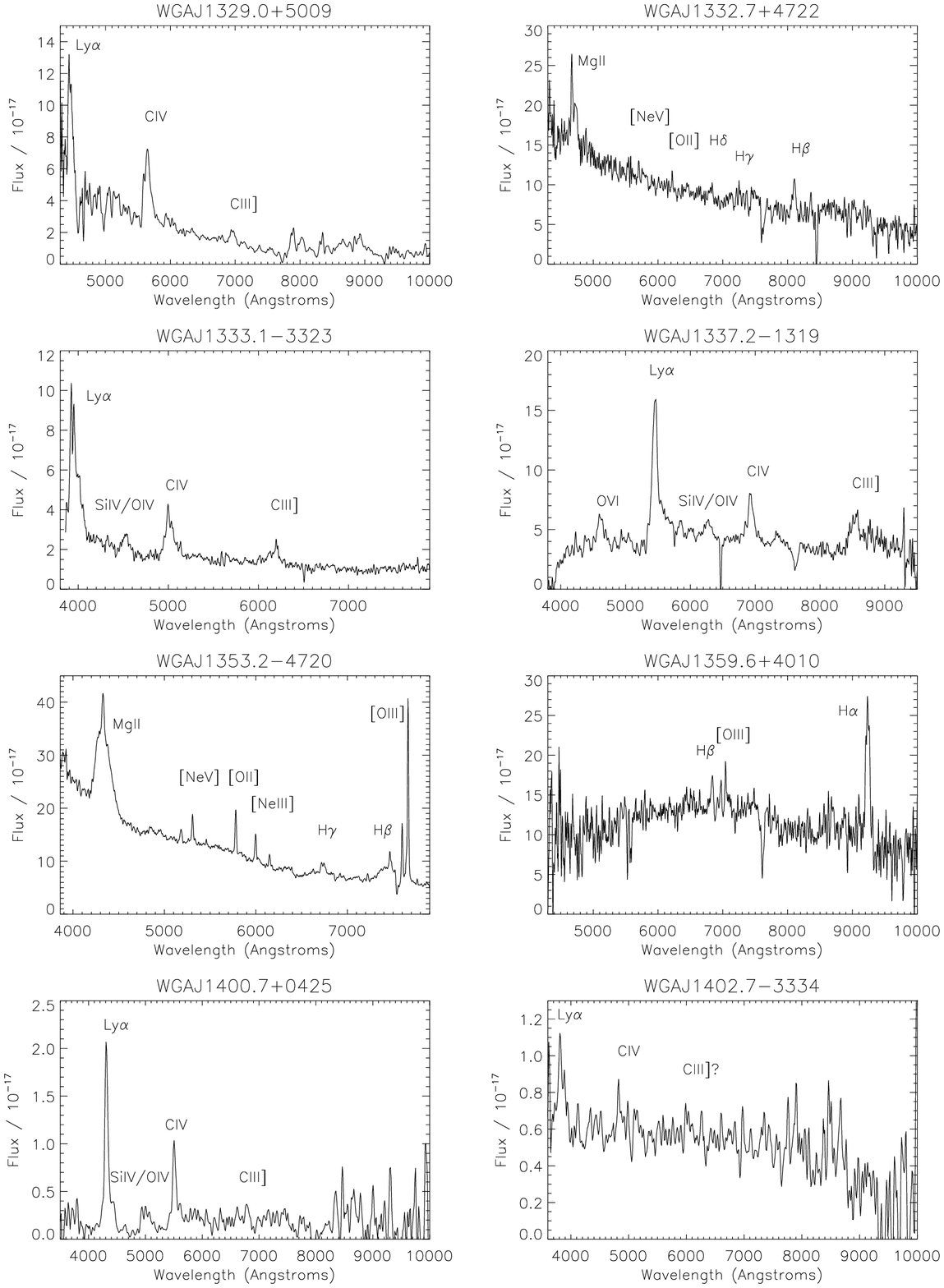,height=22.5cm}       
\end{figure*}

\begin{figure*}
\psfig{figure=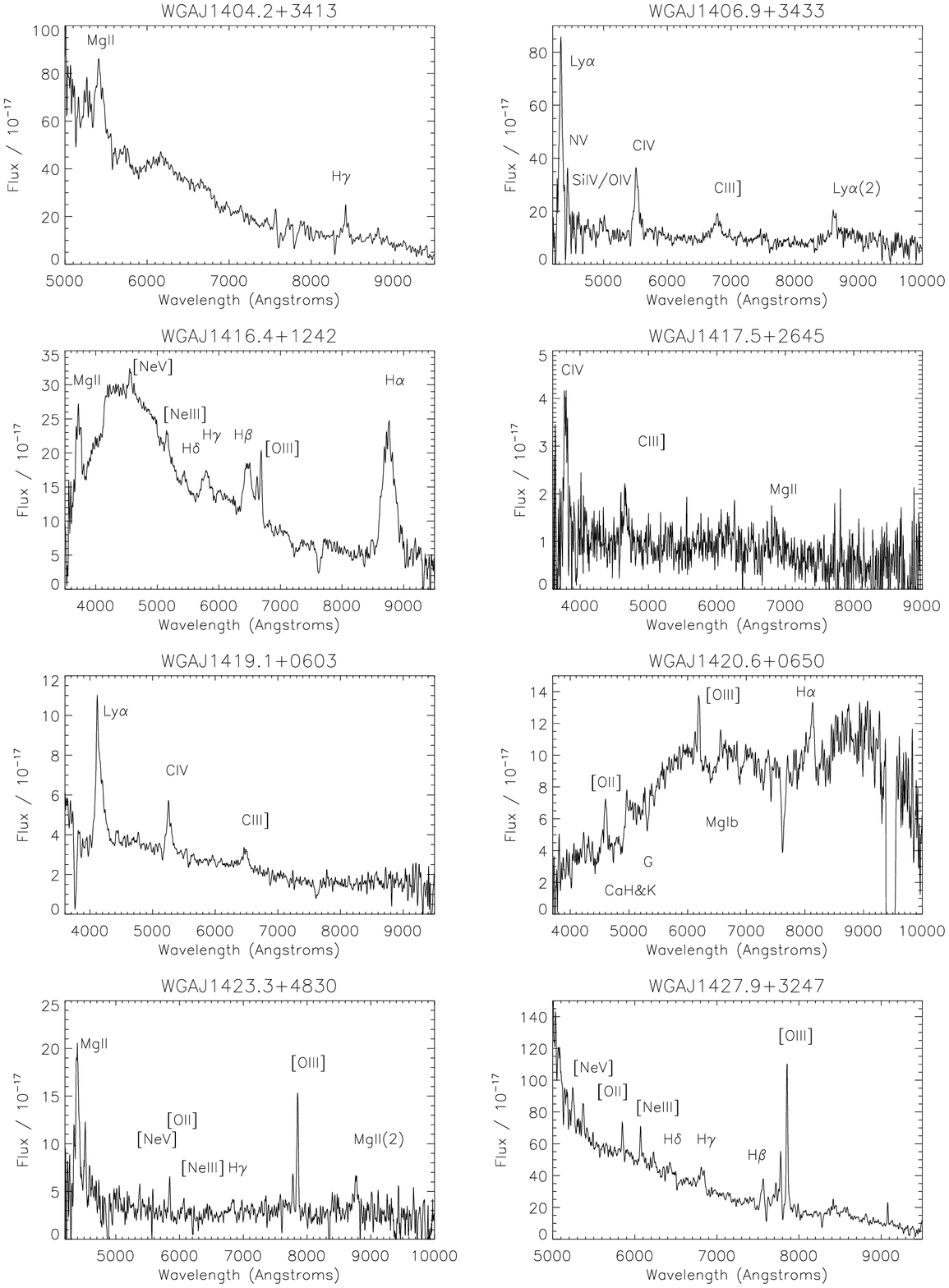,height=22.5cm}       
\end{figure*}

\begin{figure*}
\psfig{figure=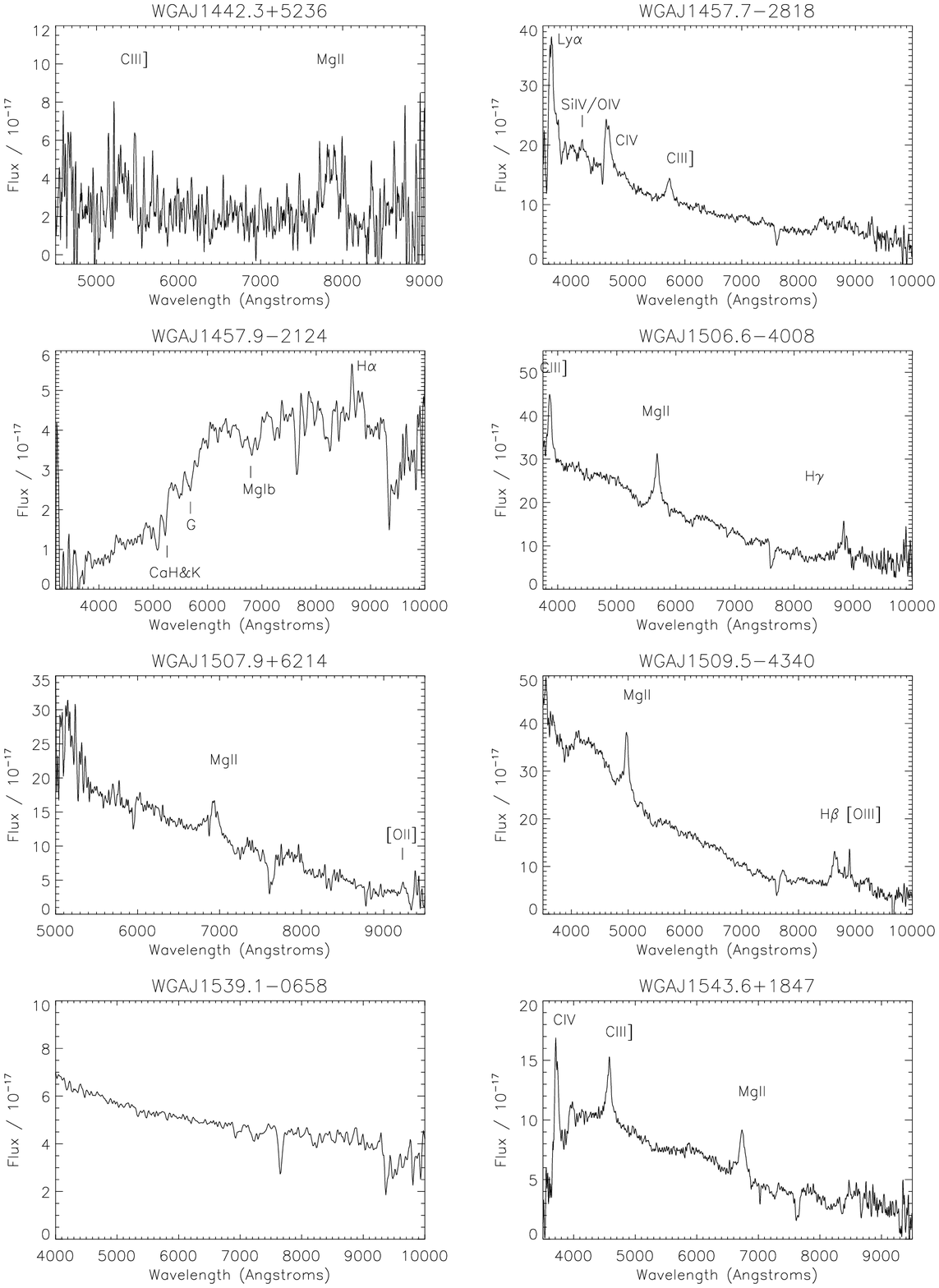,height=22.5cm}      
\end{figure*}

\begin{figure*}
\psfig{figure=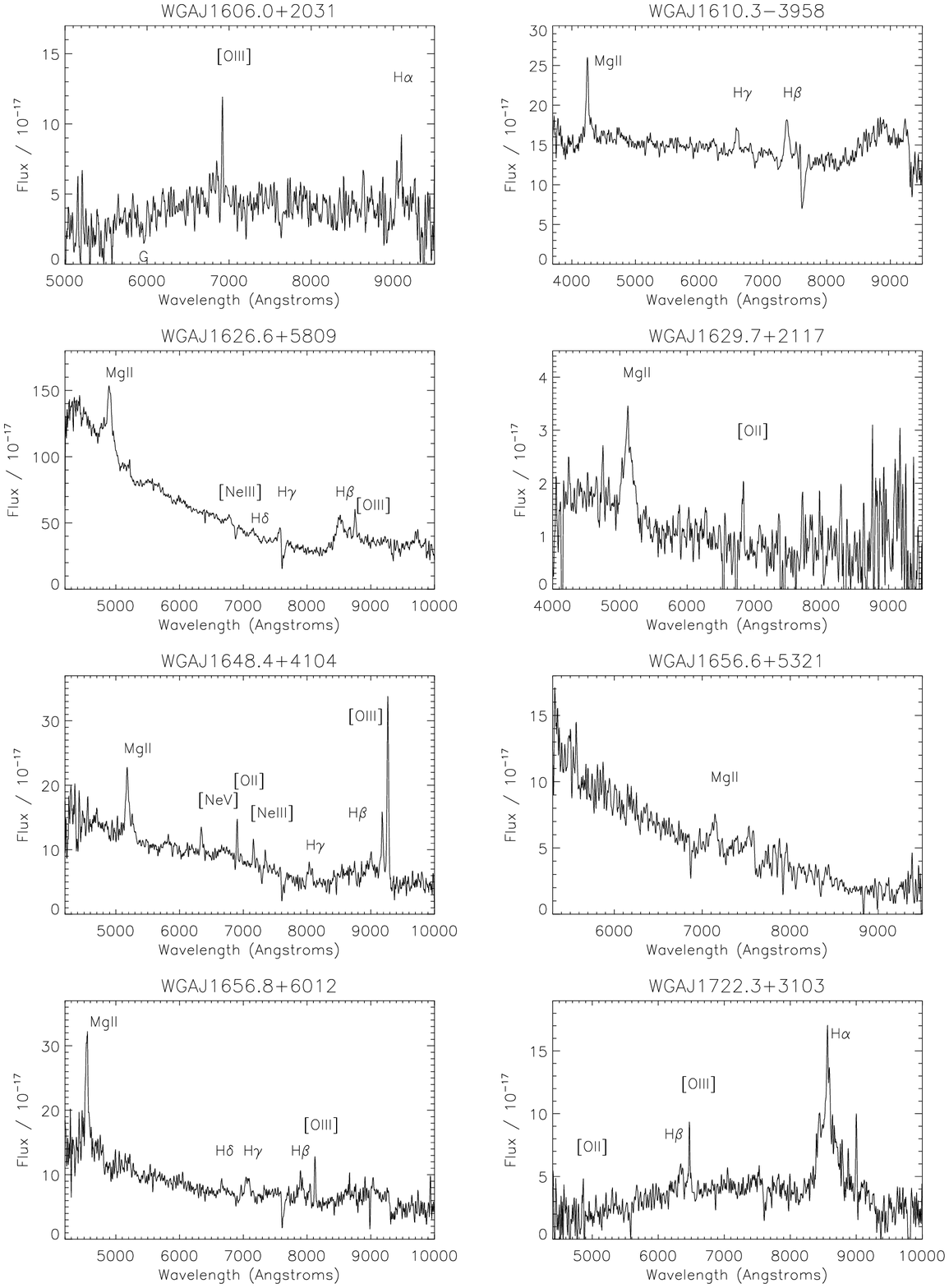,height=22.5cm}      
\end{figure*}

\begin{figure*}
\psfig{figure=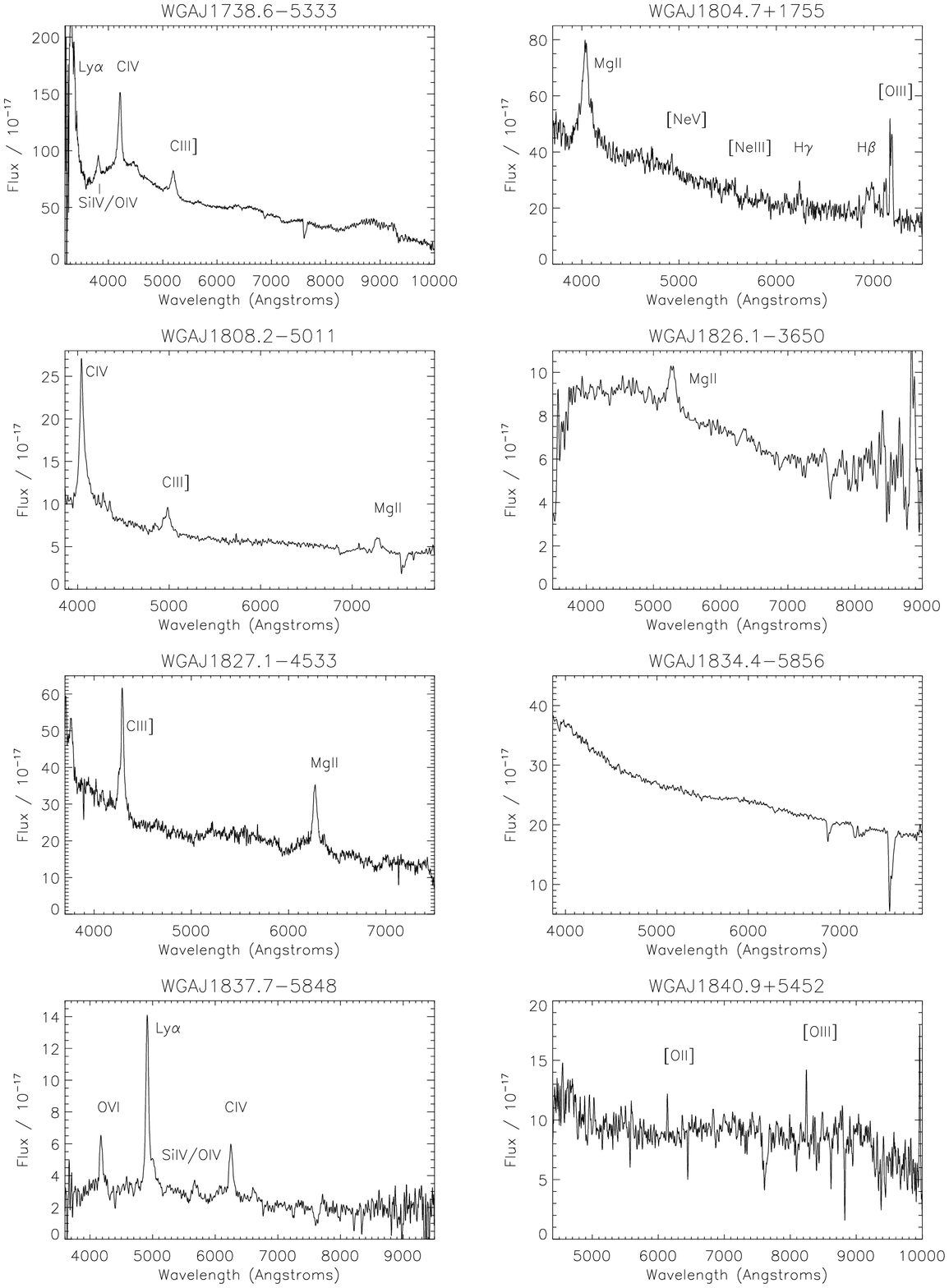,height=22.5cm}      
\end{figure*}

\begin{figure*}
\psfig{figure=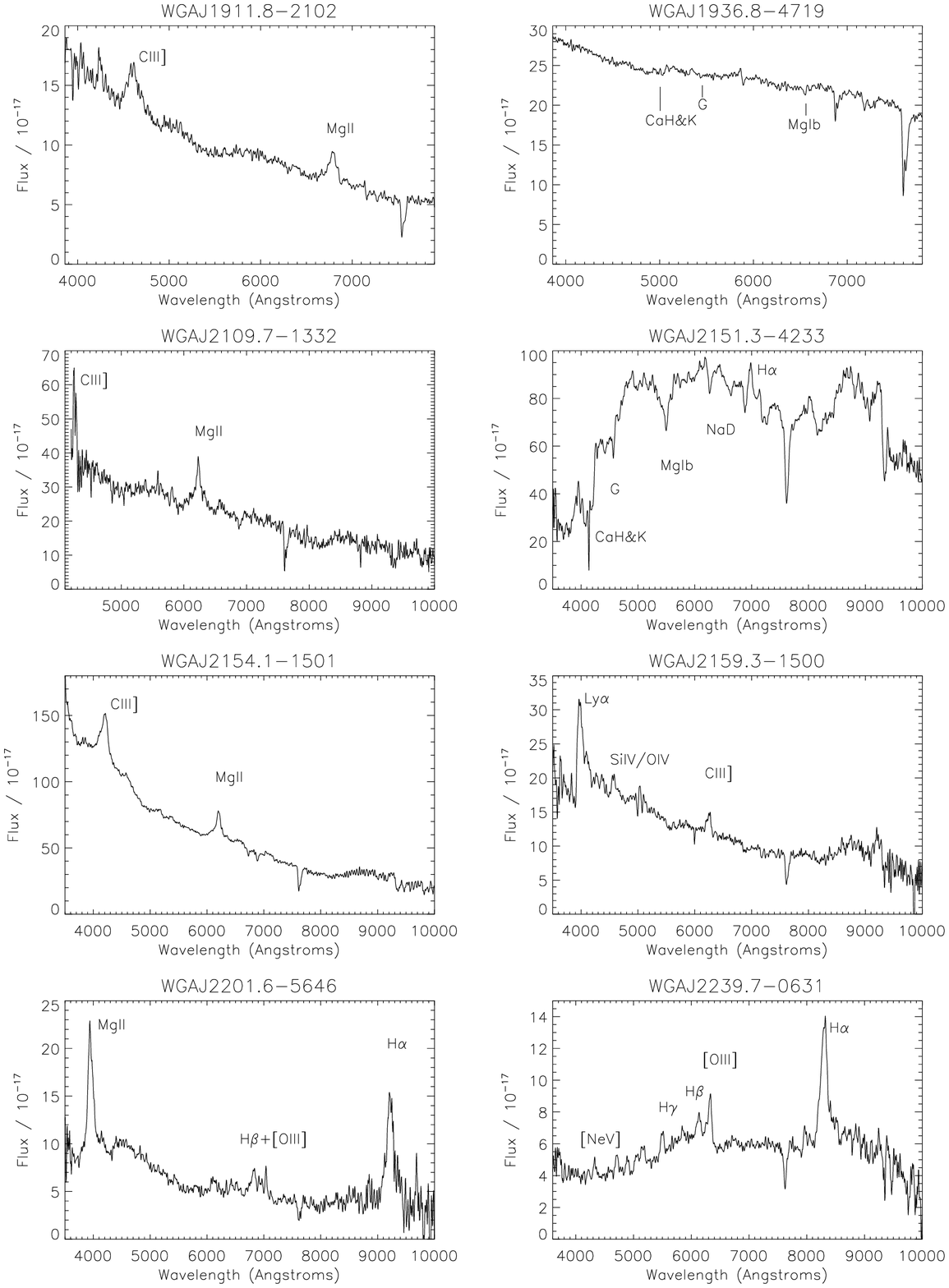,height=22.5cm}      
\end{figure*}

\begin{figure*}
\psfig{figure=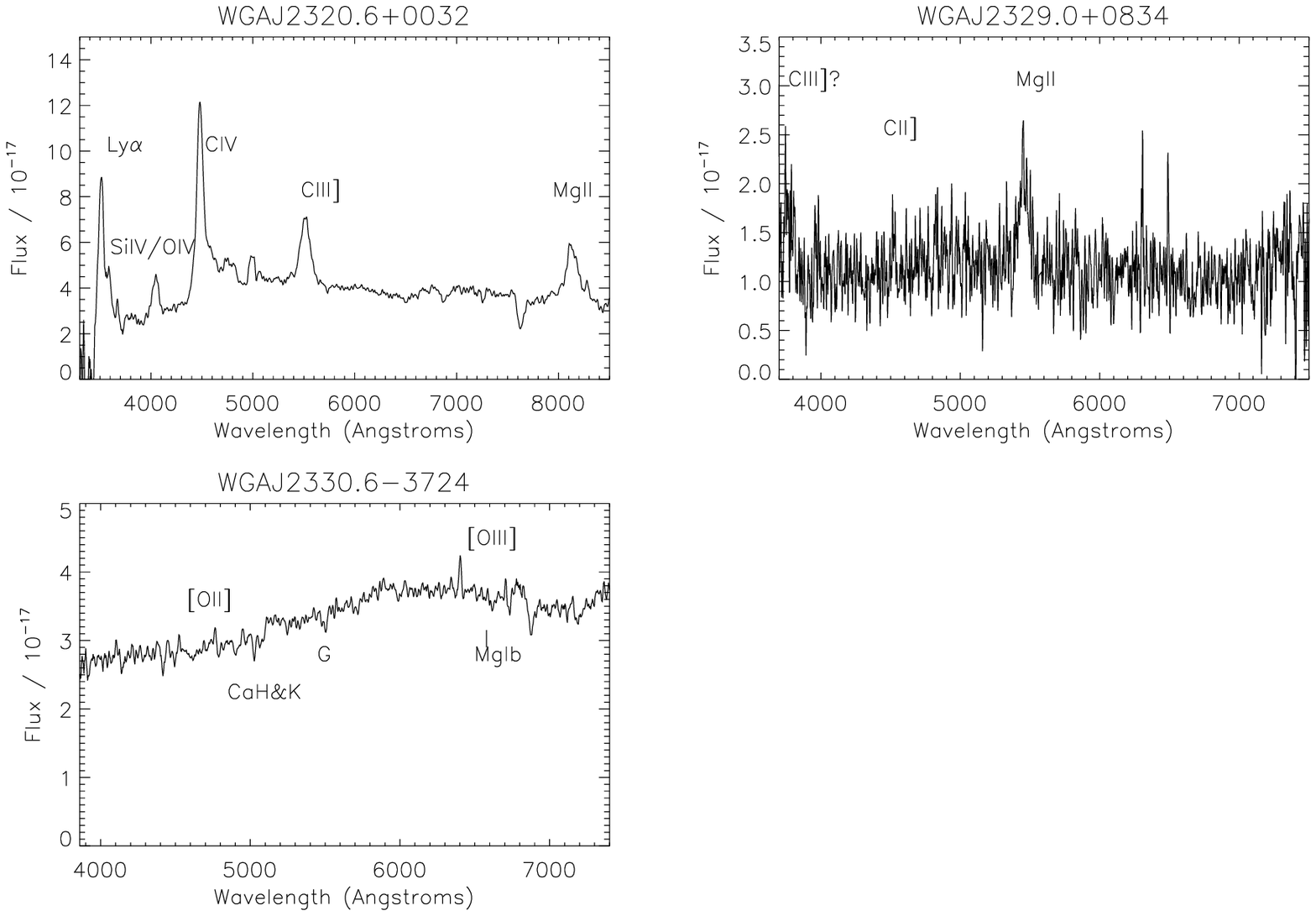,height=22.5cm}   
\end{figure*}

\end{document}